\documentclass[9.5pt,journal,compsoc]{IEEEtran}
\usepackage{graphicx}
\usepackage{booktabs}
\usepackage{amsmath}
\usepackage{amssymb}
\usepackage{amsfonts}
\usepackage{epsfig}
\usepackage{graphicx}
\usepackage{subfigure}
\usepackage{url}
\usepackage[bottom]{footmisc}
\usepackage{balance}
\usepackage{algorithmic}
\usepackage{algorithm}
\usepackage{multirow}
\usepackage{xspace}
\usepackage{capt-of}
\usepackage{pifont}

\usepackage{dsfont}
\usepackage{color}
\usepackage{cite}
\usepackage{booktabs}
\usepackage{varwidth}
\usepackage[singlelinecheck=off]{caption}
\DeclareCaptionType{copyrightbox}

\newtheorem{theor}{Theorem}

\newtheorem{lem}{Lemma}

\newtheorem{exam}{Example}
\newtheorem{problem}{Problem}
\newtheorem{obs}{Observation}

\newcommand{\blank}[1]{\hspace*{#1}}

\newcommand{\NP}{\ensuremath{\mathbf{NP}}\xspace}
\newcommand{\PTAS}{\ensuremath{\mathbf{PTAS}}\xspace}
\newcommand{\Poly}{\ensuremath{\mathbf{P}}\xspace}
\newcommand{\sharpP}{\ensuremath{\mathbf{\#P}}\xspace}
\newcommand{\bigO}{\mathcal{O}}
\DeclareMathOperator*{\argmin}{arg\,min}
\DeclareMathOperator*{\argmax}{arg\,max}

\newcommand{\spara}[1]{\smallskip\noindent{\bf #1}}

\makeatletter
\long\def\@IEEEtitleabstractindextextbox#1{\parbox{0.922\textwidth}{#1}}
\makeatother

\begin{document}
\title{Reliability Maximization in Uncertain Graphs}

\author{Xiangyu~Ke,
		Arijit~Khan,
		Mohammad~Al~Hasan,
		Rojin~Rezvansangsari
\IEEEcompsocitemizethanks{\IEEEcompsocthanksitem X. Ke and A. Khan are with NTU Singapore. \protect\\
	E-mail: \{xiangyu001, arijit.khan\}@ntu.edu.sg
\IEEEcompsocthanksitem M. A. Hasan is with Indiana University-Purdue University, IN, USA. \protect\\
	E-mail: alhasan@cs.iupui.edu
\IEEEcompsocthanksitem R. Rezvansangsari is with Sharif University of Technology. This work was done during her internship at NTU Singapore. 
}
\thanks{The research is supported by MOE Tier-1 RG83/16 and NTU M4081678.}
}
\vspace{-5mm}


\vspace{-2mm}
\IEEEtitleabstractindextext{%
\begin{abstract}
Network reliability measures the probability that a target node is reachable from a source node in an uncertain graph, i.e.,
a graph where every edge is associated with a probability of existence. In this paper, we investigate the novel
and fundamental problem of adding a small number of edges in the uncertain network for maximizing the reliability between a
given pair of nodes. We study the \NP-hardness and the approximation hardness of our problem, and design effective, scalable solutions.
Furthermore, we consider extended versions of our problem (e.g.,  multiple source and target nodes can be provided as input) to support
and demonstrate a wider family of queries and applications, including sensor network reliability maximization and social influence
maximization. Experimental results validate the effectiveness and efficiency of the proposed algorithms.
\end{abstract}

}

\maketitle

\IEEEdisplaynontitleabstractindextext
\IEEEpeerreviewmaketitle

\IEEEraisesectionheading{\section{Introduction}}
\label{sec:intr}

\IEEEPARstart{R}{ich} expressiveness of probabilistic graphs and their utility to model
the inherent uncertainty in a wide range of applications have prompted a large
number of research works on probabilistic graphs by the data management
research communities. In recent years, researchers in this community have
proposed efficient algorithms for solving several interesting problems, e.g., finding $k$-nearest neighbors~\cite{PBGK10}, answering reachability
queries~\cite{JLDW11}, and designing networks~\cite{WCWL14} --- all in an uncertain graph setting. Uncertainty
in a graph arises due to many reasons, including noisy measurements of an edge
metric~\cite{A09}, edge imputation using inference and prediction
models~\cite{AdarR07,NK03}, and explicit manipulation of edges, e.g., for
privacy purposes \cite{Boldietal12}.

In an uncertain graph setting, {\em Network Reliability} is a well-studied
problem~\cite{B86,V79}, which requires to measure the probability
that a target node is reachable from a source node. Reliability has been
widely studied in device networks, i.e., networks whose nodes are electronic
devices and the (physical) links between such devices have a probability of
failure \cite{AMG75}. More recently, the attention has been shifted to
social, communication, transportation, genomic, and logistic networks
\cite{KKT03,HP10,Ghosh2007Routing}. Applications of reliability
estimation include computing the packet delivery probability from a source to a
sink node in a wireless sensor network, measuring information diffusion
probability from an early adopter to a target customer in a social influence
network, predicting new interactions by finding all proteins that are evidently
(i.e., with high probability) reachable from a core (source) set of proteins in
a protein-interaction network, as well as estimating on-demand delivery
probability via different routes from an inventory to warehouses or customers
in a road network, among many others.

In this paper, we investigate the novel problem of adding a
small number of edges in an uncertain network for maximizing the reliability
between a given pair of nodes. We refer to such edges as {\em shortcut edges}
and the problem of identifying the best set of $k$ edges as the {\em budgeted
	reliability maximization} problem. Our problem falls under the broad category
of uncertain networks design \cite{WCWL14}, optimization \cite{Peng2015}, and
modification \cite{NSS99} problems, yet surprisingly this specific problem has not been
studied in the past.

The budgeted reliability maximization problem is critical in the context of
many physical networks, such as transportation and communication networks. In
mobile ad-hoc networks, the connectivity between sensor nodes and devices is
estimated using noisy measurements, thus leading to edges naturally associated
with a probability of existence \cite{Ghosh2007Routing}.  Road networks can be
modeled as uncertain graphs because of unexpected traffic congestion
\cite{HP10}. In these networks, creating new connections between nodes (e.g.,
building new roads, flyovers, adding Ethernet cables) is limited by physical
constraints and budget. One can introduce only $k$ new edges where $k$ is
decided based on resource constraints. Thus, our goal is to intelligently add
$k$ new edges such that the reliability between a pair of important nodes is
maximized \cite{WCWL14}. Furthermore, in social networks, finding $k$ best
shortcut edges could maximize the information diffusion probability from an
early adopter to a target customer \cite{CTPEFF16}, thus the
network host can actively recommend these links to the respective users.  In
case of protein-interaction networks, interactions are established for a
limited number of proteins through noisy and error-prone experiments---each
edge is associated with a probability accounting for the existence of the
interaction. Therefore, finding the top-$k$ shortcut edges can assist in
de-noising protein-interaction networks \cite{KRHP09}.


\vspace{-0.6mm}
\spara{Challenges and contributions.}
Unfortunately, budgeted reliability maximization problem is non-trivial. In
fact, a simpler problem to compute the exact reliability over
uncertain graphs is \sharpP-complete \cite{V79,B86}. Our thorough investigation
of the budgeted reliability maximization problem have yielded the following
theoretical results: (1) we prove that, even assuming polynomial-time
sampling methods to estimate reliability (such as, Monte Carlo sampling \cite{F86}, or
more sophisticated recursive stratified sampling \cite{LiYMJ16}), our problem
of computing a set of $k$ shortcut edges that maximizes the reliability between
two nodes remains \NP-hard; (2) the budgeted reliability maximization
problem is hard to approximate, as {\bf (i)} it does not admit any \PTAS,
and {\bf (ii)} the underlying objective function is neither
submodular, nor supermodular. The above pessimistic results are useful to
comprehend the computation challenges associated with finding even an approximate
solution to this problem, let alone an optimal one. For instance, lack of submodular (or supermodular) property prevents us from using an iterative
hill-climbing based greedy algorithm that maximizes the marginal gain at every
iteration to obtain a solution with approximation guarantees.  Moreover, a
hill-climbing algorithm would be inefficient due to repeated computation
of marginal gains for {\em all} candidate edges (which are, in fact, missing
edges in the input graph, and can be $\bigO(n^2)$ in numbers for a sparse
graph) at every iteration.

By considering the computation challenges as we have discussed above, in this
paper we propose a practical algorithm for budgeted reliability maximization
problem.  Our proposed solution systematically minimizes the search space by
only considering missing edges between nodes that have reasonably high
reliability from the source node and to the target node.  Next, we extract
several highly-reliable paths between source and target nodes, after including
those limited number of candidate edges in the input graph. This is motivated
by the observation that what really matters in computing the reliability
between two nodes is the set of paths connecting source to target, not the
individual edges in the graph \cite{KBGN18,CWW10,KZK16}. Our algorithm
then iteratively selects these paths so as to achieve maximum improvement in
reliability while satisfying the constraint on the number of new edges ($k$) to
be added.

We also consider a {\em restricted} version of our problem, which approximates
the reliability by considering {\em only} the most reliable path between the
source and the target node \cite{KZK16,CWW10}. We prove that improving the
probability of the most reliable path can be solved exactly in polynomial time,
which yields an efficient algorithm for the restricted version of our
problem.  Finally, we focus on {\em generalizations} where
multiple source and target nodes can be provided as input, thus opening the
stage to a wider family of queries and applications, e.g., network modification
for targeted influence maximization \cite{CTPEFF16,KKC18}.

The main contributions of this paper are as follows.
\vspace{-2mm}
\begin{itemize}
	\setlength\itemsep{0.01em}
	\item We study the novel and fundamental problem of maximizing the reliability between a given pair of nodes by adding a small number of edges in an uncertain graph. Our problem is \NP-hard,
	and is also hard to approximate, even when polynomial-time reliability estimation is employed (\S\ref{sec:preliminaries}).
	\item We design effective and efficient solutions for our problem. Our algorithms first apply reliability-based search space elimination, then fill the remaining graph with missing edges, and finally select the
	top-$k$ edges to add based on most reliable paths (\S\ref{sec:proposed}).
	\item We further consider a restricted and one extended version of our problem to support a wider family of queries. In the restricted version, the reliability is estimated only by the most reliable path, thus it
	can be solved 
	in polynomial-time. In the extended version, multiple sources and targets can be provided as input. The proposed algorithms are generalized to multiple-source-target case. (\S\ref{sec:relpath} and \S\ref{sec:multi}).
	\item We conduct a thorough experimental evaluation with several real-world and synthetic graphs
	to demonstrate the effectiveness, efficiency, and scalability of our algorithms, and illustrate the
	usefulness of our problem in critical applications such as sensor network reliability maximization and influence maximization in social networks (\S\ref{sec:exp}).
\end{itemize} 

\vspace{-3mm}
\section{Preliminaries}
\label{sec:preliminaries}

\begin{table}[tb!]
	\centering
	\caption{Table of notations (used in \S~\ref{sec:preliminaries}: Preliminaries).}
	\vspace{-2mm}
	\begin{tabular} {c||l}
		\hline
		{\bf Notation} & {\bf Description}  \\
		\hline \hline
		$\mathcal{G}$ & input uncertain graph, represented by  triple $(V,E,p)$ \\ \hline
		$V$ & set of $n$ nodes \\ \hline
		$E$ & set of $m$ directed edges, $E\subseteq V\times V$ \\ \hline
		$p(e)$ & probability that the edge $e\in E$ exist, $p(e)\in [0,1]$ \\ \hline
		\multirow{2}{*}{$G$} & deterministic graph instance of $\mathcal{G}$, represented by \\ &$(V,E_G)$ \\ \hline
		\multirow{2}{*}{$E_G$} & set of edges in $G$, $E_G \subseteq E$, obtained by independent\\ 
		&  sampling \\ \hline
		$Pr(G)$ & probability of a graph instance $G$ being observed \\ \hline
		\multirow{2}{*}{$I_G(s,t)$} & indicator function, which takes value 1 if there exists\\ & a path from $s$ to $t$ in $G$, and 0 otherwise \\ \hline
		\multirow{2}{*}{$R(s,t,\mathcal{G})$} & the $s$-$t$ reliability in uncertain graph $\mathcal{G}$, which is the  \\ & probability that $t$ is reachable from $s$ in  $\mathcal{G}$\\ \hline
		$k$ & budget on the number of new edges \\ \hline
		$\zeta$ & probablity threshold for new edges \\ \hline
		$h$ & distance threshold for new edges \\ \hline
		\hline
	\end{tabular}
	\vspace{1mm}
	\label{tab:notation_1}
	\vspace{-5mm}
\end{table}

\subsection{Problem Formulation}
\label{sec:prob}
\vspace{-1mm}
An uncertain graph $\mathcal{G}$ is a triple $(V,E,p)$, where $V$ is a set of $n$ nodes, $E \subseteq V \times V$ is a set of $m$ directed edges, and $p(e)\in[0,1]$ is the probability that the edge $e\in E$ exists. Following bulk of the literature on uncertain graphs \cite{V79,B86,JLDW11,KBGN18,PBGK10,KYC18}, we assume that edge probabilities are independent of each other. Therefore, we employ the well-established {\em possible world} semantics: The uncertain graph $\mathcal{G}$ yields $2^m$ deterministic graphs $G\sqsubseteq \mathcal{G}$. Each possible world $G=(V,E_G)$ is a certain instance of the uncertain graph $\mathcal{G}$, where $E_G\subseteq E$ and is obtained by independent sampling of the edges. Its probability of being observed is given as:
\begin{align}
\vspace{-3mm}
Pr(G)=\prod_{e\in E_G}p(e)\prod_{e\in E \setminus E_G}(1-p(e))
\vspace{-3mm}
\end{align}
Given a source node $s\in V$, a target node $t\in V$, the {\em reliability} $R(s,t,\mathcal{G})$, also known as
the $s$-$t$ reliability, is defined as the probability that $t$ is reachable from $s$ in $\mathcal{G}$.
Formally, for a possible graph $G \sqsubseteq \mathcal{G}$, let $I_G(s,t)$ be an indicator function taking the value $1$ if there exists a path from $s$ to $t$ in $G$, and $0$ otherwise. $R(s,t,\mathcal{G})$ is computed as follows.
\begin{align}
\vspace{-3mm}
R(s,t,\mathcal{G})=R(s,t,(V,E,p))=\sum_{G \sqsubseteq \mathcal{G}} [I_G(s,t)\times Pr(G)]
\label{eq:r}
\vspace{-4mm}
\end{align}
The problem that we study in this work is stated below.
\begin{problem}
[Single-source-target budgeted reliability maximization] Given an uncertain graph $\mathcal{G}=(V,E,p)$, a source node $s\in V$, a target node $t\in V$, a probability threshold $\zeta\in(0,1]$,
and a small positive integer $k$,
find the top-$k$ edges to add in $\mathcal{G}$,
each with probability $p(e)=\zeta$, so that the reliability from $s$ to $t$ is maximized.
\vspace{-1mm}
\begin{align}
& E^*= \argmax_{E_1\subseteq V\times V\setminus E }R \left(s,t,\left(V,E\cup E_1,p\right)\right) \nonumber  \\
&\text{s. t.} \qquad |E_1|=k; \qquad \text{and} \quad p(e)=\zeta \quad \forall e \in E_1
\vspace{-3mm}
\end{align}
\label{prob:single-st}
\vspace{-3mm}
\end{problem}
\vspace{-5mm}

For simplicity, we adopt a fixed probability threshold $\zeta$ on new edges.
The intuition is that when establishing a new edge, generally we consider a
connection with the best/average possible reliability, e.g., an Ethernet cable
with the highest reliability in case of LAN, or the average link reliability of
sensor edges as we have used in the discussed case study. However, if the user provides probability values for the missing edges as part of input,
our proposed algorithm (\S\ref{sec:proposed}) will work smoothly as it can
simply use those values instead of $\zeta$ when finding the most reliable paths
(see our experimental evaluation in Table 16,
\S\ref{sec:single_exp}).

\vspace{-0.6mm}
\spara{Remarks.} Due to various physical and resource constraints, in practice, it might not be possible to consider
all missing edges in the input graph as candidates for our problem. In a social network it is often realistic to
recommend new connections between users who are within 2-3 hops. In a communication network,
a new edge can be added only if the two nodes are within a certain geographical distance. While we analyze the complexity of our problem
and develop algorithms for the generalized case, that is, all missing edges can potentially be candidate edges, in our solution
as well as in experiments we provision for a threshold distance $h$: Two nodes can be added by a new edge only if they are
within $h$-hops away. Note that {\bf (1)} when $h$ is the diameter of the graph (i.e., maximum shortest-path distance
between any pair of nodes), this is essentially equivalent to the generalized case.
{\bf (2)} Smaller values of $h$ reduces search space, thereby improving efficiency. In our experiments,
we analyze scalability of our methods for different values of $h$.

\vspace{-2mm}
\subsection{Hardness of the Problem}
\label{sec:hard}
\vspace{-1mm}
Problem~\ref{prob:single-st} depends on reliability computation in uncertain graphs, which is \sharpP-complete \cite{V79,B86}.
Thus, single-source-target budgeted reliability maximization problem is hard as well.
However, as reliability can be {\em estimated} in polynomial time via Monte Carlo (MC) sampling \cite{F86},
or more sophisticated recursive stratified sampling \cite{LiYMJ16}, the key question
is whether Problem~\ref{prob:single-st} remains hard even if polynomial-time reliability estimation methods are employed.
Due to combinatorial nature of our problem, and assuming $\bigO(n^2)$ missing edges in a sparse graph,
one can design an {\em exact} solution that compares the $s$-$t$ reliability gain for $\binom{n^2}{k}$ possible
ways of adding $k$ new edges, and then reports the best one. However, this is clearly infeasible
for large networks. We, in fact, prove that our problem is \NP-hard, and it does not admit any \PTAS.
Moreover, Problem~\ref{prob:single-st} is neither submodular, nor supermodular for inclusion of edges.
\begin{figure}[tb!]
	\footnotesize
	\vspace{-1.5mm}
	\parbox{.58\linewidth}{
		\centering
		\includegraphics[scale=0.22]{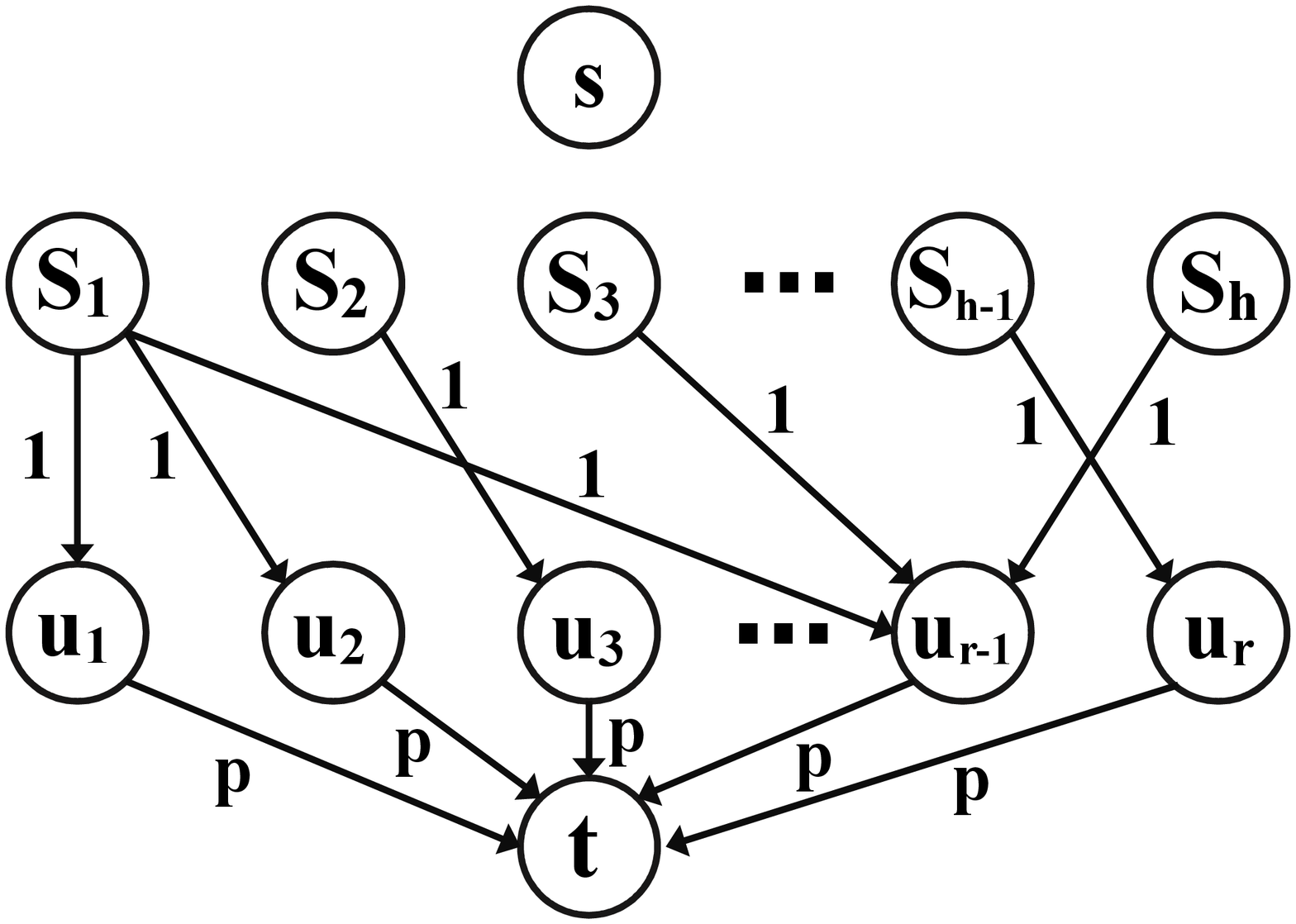}
		\vspace{-3mm}
		\caption{Reduction from MAX $k$-COVER to single $s$-$t$ budgeted reliability maximization problem}
		\label{fig:reduction}
	}
	\hfill
	\parbox{.36\linewidth}{
		\centering
		\vspace{5mm}
		\includegraphics[scale=0.52]{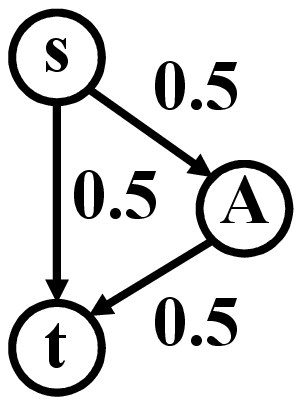}
		\vspace{0.5mm}
		\caption{Example for non-submodularity, non-supermodularity}
		\label{fig:sub}
	}
	\vspace{-5.5mm}
\end{figure}
\vspace{-0.5mm}
\begin{theor}
Problem~\ref{prob:single-st} is \NP-hard in the number of newly added edges, $k$.
\label{th:np}
\end{theor}
\vspace{-1.5mm}
\begin{IEEEproof}
We prove \NP-hardness by a reduction from the MAX $k$-COVER problem, which
is \NP-hard. In MAX $k$-COVER, there is a collection of subsets $S=\{S_1, S_2, ..., S_h\}$ of a
ground set $U=\{u_1, u_2, ..., u_r\}$, where $S_i \subseteq U$ for all $i\in [1...h]$.
The objective is to find a subset $S^*\subset S$ of size $k$ such that maximum number of
elements in $U$ can be covered by $S^*$, i.e., so as to maximize $|\bigcup_{S_i\in S^*}S_i|$.
For an instance of MAX $k$-COVER, we construct an
instance of our Problem~\ref{prob:single-st} in polynomial time as follows (Figure~\ref{fig:reduction}).

We create an uncertain graph $\mathcal{G}$ with a source node $s$ and a target node $t$.
For each element $u_i$ in $U$, we add a node in $\mathcal{G}$. Each $u_i$ is connected to $t$ by an
edge with probability $p$, such that $0<p<1$. Then, we also add each $S_i$ in $S$ as a node of $\mathcal{G}$.
Node $S_i$ is connected to node $u_j$ with probability 1 if and only if $u_j\in S_i$.
All other edges in $\mathcal{G}$ except those from $s$ to all $S_i$ have probability 0.
	
Thus, the candidate set of edges to add in $\mathcal{G}$ for maximizing reliability
from $s$ to $t$ are those edges from $s$ to all $S_i$. Without them, there is no path from $s$ to $t$ with non-zero probability.
Let $\zeta=1$, after $k$ of these edges are selected, $q$ out of $r$ elements in $U$ are now reachable from $s$,
then the $s$-$t$ reliability $=1-(1-p)^q$, which monotonically increases with larger $q$.
This implies that Problem~\ref{prob:single-st} and MAX $k$-COVER are equivalent here.
If there exists a polynomial time solution for Problem~\ref{prob:single-st}, the MAX $k$-COVER can be solved in polynomial time too.
The theorem follows.
\end{IEEEproof}

Moreover, Problem~\ref{prob:single-st} is also hard to approximate.
\vspace{-0.5mm}
\begin{theor}
Problem~\ref{prob:single-st} does not admit any \PTAS, unless \Poly=\NP.
\label{th:ptas}
\end{theor}
\vspace{-1.5mm}
\begin{IEEEproof}
A problem is said to admit a {\em Polynomial Time Approximation Scheme} (\PTAS) if the problem admits
a polynomial-time constant-factor approximation algorithm
for every constant $\beta \in (0, 1)$. We prove the theorem by showing
that one can find at least one value of $\beta$ such that, if a
$\beta$-approximation algorithm for Problem~\ref{prob:single-st} exists,
then we can solve the well-known SET COVER problem in polynomial time.
Since SET COVER is \NP-hard, this can happen only if \Poly = \NP.

In SET COVER, there is a collection of subsets $S=\{S_1, S_2,$ $\ldots,S_h\}$ of a
ground set $U=\{u_1, u_2, ..., u_r\}$, where $S_i \subseteq U$ for all $i\in [1...h]$.
The decision version of SET COVER asks if there is a subset $S^*\subset S$ of size $k$ such
that {\em all} elements in $U$ can be covered by $S^*$.

Given an instance of SET COVER, we construct an instance
of our problem in polynomial time by following the same method as in {\bf NP}-hardness proof
(Figure~\ref{fig:reduction}). In the SET COVER instance, if there is a solution with $k$
subsets, then the optimal solution {\sf OPT} of our problem will add $k$ edges such that the
$s$-$t$ reliability after edge addition is: $1-(1-p)^r$. This is because $|U|=r$.
In contrast, if no solution with
$k$ subsets exists for SET COVER, then {\sf OPT} will produce $s$-$t$ reliability
at most: $1-(1-p)^{r-1}$ (because at least one of $u_j$ would not be covered).

Let there be a polynomial-time $\beta$-approximation
algorithm, {\sf Approx}, for Problem~\ref{prob:single-st}, such that $0 < \beta < 1$.
According to the definition of approximation ratio,
{\sf Approx} will produce $s$-$t$ reliability at least $\beta$ times to that produced by {\sf OPT}.
Now, let us consider the inequality: $1-(1-p)^{r-1}<\beta[1-(1-p)^r]$.
If this inequality has a solution for some values of $\beta$ and $p$, then by simply
running {\sf Approx} on our instance of Problem~\ref{prob:single-st}, and checking the $s$-$t$ reliability of the
solution returned by {\sf Approx}, one can answer SET COVER
in polynomial time: a solution to SET COVER exists iff the
solution given by {\sf Approx} has $s$-$t$ reliability $\ge \beta[1-(1-p)^r]$.
Thus, to prove the theorem, we require to show that a solution
to that inequality exists.

Our inequality has a solution iff $\beta > \frac{1-(1-p)^{r-1}}{1-(1-p)^r}$.
One can verify that $\frac{1-(1-p)^{r-1}}{1-(1-p)^r}< 1$, for all $r \ge 1$ and $p > 0$.
This implies that there will always be a value of $\beta \in (0,1)$ and $p$ for which $\beta > \frac{1-(1-p)^{r-1}}{1-(1-p)^r}$
is satisfied, regardless of $r$. Hence, there exists at least one value of $\beta$ such that the inequality
$[1-(1-p)^{r-1}] < \beta [1-(1-p)^r]$ has a solution, and, based on the above argument, such that no $\beta$-approximation
algorithm for Problem \ref{prob:single-st} can exist. The theorem follows.
\end{IEEEproof}

We further show that neither submodularity nor supermodularity holds for the objective function of
Problem~\ref{prob:single-st}, and demonstrate with the following counter example. Therefore, standard
greedy hill-climbing algorithms do not directly come with
approximation guarantees for Problem~\ref{prob:single-st}.
\begin{lem}
	The objective function of Problem~\ref{prob:single-st} is neither submodular, nor supermodular w.r.t inclusion of edges.
	\vspace{-1mm}
\end{lem}
%

For any set $X\subseteq Y$ and all elements $x\notin Y$, a set function $f$ is submodular if $f(X\cup\{x\})-f(X)\geq f(Y\cup \{x\})-f(Y)$.
For supermodularity, the inequality is reversed.

Let us consider the example in Figure~\ref{fig:sub}: $s$ is the source node and $t$ is the target nodes.
Assume the node set $V=\{s, A, t\}$.
Let $X=\{st\}$, $Y=\{st, sA\}$ be two edge sets.
We have $R\left(s,t,\left(V,X,p\right)\right)=R\left(s,t,\left(V,Y,p\right)\right)=0.5$.
We find that $R\left(s,t,\left(V,X\cup\{At\},p\right)\right)=0.5$, $R\left(s,t,\left(V,Y\cup\{At\},p\right)\right)=1-(1-0.5)[1-0.5^2]=0.625$.
Clearly, submodularity does not hold in this example.

Next, considering $X'=\{sA\}$, $Y'=\{sA,st\}$, we have $R\left(s,t,\left(V,X',p\right)\right)=0$ and $R\left(s,t,\left(V,Y',p\right)\right)=0.5$.
Then, $R\left(s,t,\left( V,X' \cup \{At\},p \right) \right)=0.25$, $R\left(s,t,\left(V,Y'\cup \{At\},p\right)\right)\\=0.625$, thus supermodularity also does not hold.

\vspace{-4mm}
\subsection{Characterization of the Problem}
\label{sec:character}
\vspace{-1mm}
We next show
that the optimal solution to our problem varies based on most input parameters, even if the
other set of input parameters remains the same, thereby making it non-trivial to utilize pre-existing solutions
of past queries, as well as indexing-based or incremental methods.

\begin{obs}
The optimal solution for Problem~\ref{prob:single-st} may vary with different input probability threshold $\zeta$.
\label{ob:zeta}
\end{obs}
\vspace{-2mm}
\begin{obs}
The optimal solution for Problem~\ref{prob:single-st} may vary when the edge probabilities in the original graph change.
\label{ob:prob}
\end{obs}
\vspace{-6mm}
\begin{obs}
When $k_1<k_2$, the optimal solution for Problem~\ref{prob:single-st} with $k_1$ may not be a subset of that with $k_2$.
\vspace{-1mm}
\label{ob:sub}
\end{obs}
\begin{table}[t!]
	\vspace{-1mm}
	\begin{minipage}[b]{0.7\linewidth}
		\centering
		\begin{scriptsize}
			\captionof{table}{Reliability gains of three possible solutions for the example in Figure \ref{fig:char} under different setting.}
			\vspace{-3mm}
			\label{tab:char}
			\begin{tabular} {cc||ccc}
				\hline
				\multirow{2}{*}{$\alpha$} & \multirow{2}{*}{$\zeta$} & \multicolumn{3}{c}{\bf Reliability}  \\ \cline{3-5}
				& & $\{sA,sB\}$ & $\{sA,Bt\}$ & $\{sB,Bt\}$ \\ \hline \hline
				0.5 & 0.7 & 0.403 & 0.473 & {\bf 0.543}  \\
				0.5 & 0.3 & {\bf 0.203} & 0.173 & 0.143 \\
				0.9 & 0.7 & {\bf 0.800} & 0.674 & 0.660  \\ \hline
			\end{tabular}		
		\end{scriptsize}
		\vspace{-5mm}
	\end{minipage}
	\hfill
	\begin{minipage}[b]{0.28\linewidth}
		\centering
		\vspace{-3mm}
		\includegraphics[scale=0.395]{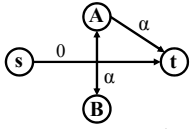}
		\vspace{-7mm}
		\captionof{figure}{Example for the problem characterization}
		\label{fig:char}
		\vspace{-5mm}
	\end{minipage}
	\vspace{1mm}
\end{table}

All these three observations can be demonstrated with the example given in Figure \ref{fig:char}, as follows.
\begin{exam}
In Figure \ref{fig:char}, there are edges $AB$ and $At$, both with probability $\alpha$ ($0<\alpha<1$), in this graph.
And the edge directly connecting $s$ and $t$ can not exist, e.g., no direct flight can be established between two airports if they
are too far away. Clearly, original reliability between $s$ and $t$ is 0. $\{sA,sB,Bt\}$ is the candidate set of
edges to add for improving the $s$-$t$ reliability.
	
If budget $k=1$, $\{sA\}$ is always the optimal solution. Its reliability is $\alpha\zeta$, which is larger than both
$\alpha^2\zeta$ for solution $\{sB\}$ and 0 for solution $\{Bt\}$.
	
If budget $k=2$, there are 3 possible solutions, $\{sA,sB\}$, $\{sA,Bt\}$, and $\{sB,Bt\}$. The reliability between $s$ and $t$ after adding them can be calculated as follows:
\vspace{-2mm}
\begin{align}
&R(s,t,(V,E\cup\{sA,sB\},p))=[1-(1-\zeta)(1-\alpha\cdot\zeta)]\cdot\alpha \nonumber \\
& R(s,t,(V,E\cup\{sA,Bt\},p))=\zeta\cdot[1-(1-\alpha)(1-\alpha\cdot\zeta)] \nonumber \\
& R(s,t,(V,E\cup\{sB,Bt\},p))=\zeta\cdot[1-(1-\zeta)(1-\alpha^2)] \nonumber
\vspace{-2mm}
\end{align}
Table \ref{tab:char} presents the reliability of these solutions with different $\alpha$ and $\zeta$. Clearly, rows 1 and 2 have same $\alpha$ and different $\zeta$, and their optimal solutions are different:
$\{sA,sB\}$ and $\{sB,Bt\}$, respectively. This confirms our Observation \ref{ob:zeta}. Similarly, we have same $\zeta$ but different $\alpha$ in rows 1 and 3, and obtain different optimal solutions. Therefore, we draw Observation \ref{ob:prob}. Moreover, $\{sA\}$, the optimal solution when $k=1$, is not a subset of the optimal solution $\{sB,Bt\}$ when $k=2$ if $\alpha=0.5$, $\zeta=0.7$, which implies our Observation \ref{ob:sub}.
\vspace{-1mm}
\label{ex:char}
\end{exam}
Finally, we conclude this section with an interesting observation below: The direct edge $st$, if missing in the input graph, will always be in the top-$k$
optimal solution. In other words, when the direct $st$ edge is missing and if it can be added, for the top-$1$ solution, adding the direct $st$
edge is the best solution.
\vspace{-0.5mm}
\begin{obs}
	If the direct edge from $s$ to $t$, $st$, is missing in the input graph, and is allowed to be added, $st$ will always be included in the top-$k$ optimal solution.
	\vspace{-1mm}
	\label{ob:st}
\end{obs}
\begin{IEEEproof}
Let $G$ be a possible world (i.e., deterministic graph) of the original uncertain graph $\mathcal{G}$. Following Equation~\ref{eq:r},
the $s$-$t$ reliability is calculated as: $\sum_{G \sqsubseteq \mathcal{G}} [I_G(s,t)\times Pr(G)]$. After adding $k$ missing edges,
$G$ will partition into $2^k$ new possible worlds: $\{G_1,G_2,$ $\ldots,G_{2^k}\}$. $Pr(G)=\sum_{i=0}^{2^k}Pr(G_i)$. Clearly, when $t$ is reachable
from $s$ in $G$, it will still be reachable from $s$ in each of $\{G_1,G_2,...,G_{2^k}\}$, thus $I_{G_i}(s,t)$ will continue to be 1.
Therefore, we only investigate those $G$ containing no path from $s$ to $t$, where the $s$-$t$ reliability can be improved with new edges.

Suppose $\{e_1,e_2,...,e_k\}$ is an optimal solution without $st$. For any $G_i$ in $\{G_1,G_2,...,G_{2^k}\}$ obtained from some $G$,
where $t$ was originally not reachable from $s$, we consider another solution by replacing $e_j$ ($1\leq j \leq k$) with $st$.
{\bf (1)} If $e_j$ exists in $G_i$, $I_{G_i}$ may or may not be 1. However, when replacing $e_j$ with $st$,
$I_{G_i}$ will always return 1, and improve the reliability; {\bf (2)} If $e_j$ is absent in $G_i$, the value of $I_{G_i}$ depends only on other
edges in the solution set. Replacing $e_j$ with $st$ will not impact the reliability. Therefore, replacing $e_{j}$ with $st$ will result in a new solution which has reliability gain at least as large as the earlier one. This implies that $st$, if allowed, can always be added in the optimal solution.
\end{IEEEproof}

\vspace{-2mm}
\section{Baseline Methods}
\label{sec:algo}
\vspace{-1mm}
In this section, we first present several baseline methods, that are straightforward,
and demonstrate how they suffer from both effectiveness and efficiency issues.
The discussions will be instrumental in developing a more accurate and scalable
solution in \S~\ref{sec:relpath} and \S~\ref{sec:proposed}.

\vspace{-3mm}
\subsection{Individual Top-$k$ Method}
\label{sec:enu}
\vspace{-1mm}
In the most straightforward approach, we consider every candidate edge
one by one, check the reliability gain due to its addition in the input graph
with probability $\zeta$, and select the top-$k$ edges with highest
individual reliability gains.

\vspace{-0.6mm}
\spara{Time complexity.} The reliability can be estimated in polynomial time via
Monte Carlo (MC) sampling. It samples $Z$ deterministic
graphs from the input uncertain graph, and estimates the reliability of an $s$-$t$ pair
as ratio of samples 
in which the target is reachable from the source. The reachability in a deterministic graph can be
evaluated via breadth first search (BFS) in time $\bigO(n+m)$, where $n$ and $m$ denote
the number of nodes and edges in the input graph, respectively. Thus, the time complexity
of MC sampling for each newly added edge is $\bigO(Z(n+m))$.
Since real-world networks are generally sparse, the number of candidate edges is nearly $\bigO(n^2)$.
Therefore, the overall complexity of individual top-$k$ baseline is: $\bigO(n^2Z(n+m)+n^2\log k)$,
where the last term is due to top-$k$ search.

\vspace{-0.6mm}
\spara{Shortcomings.} {\bf (1)} To achieve reasonable accuracy, MC sampling requires around thousands of samples \cite{JLDW11,KKT03}.
Performing this for $\bigO(n^2)$ times is not scalable for large graphs.
{\bf(2)} Once an edge is added into the input graph, the reliability gain of adding other candidate edges may change.
Hence, selecting the top-$k$ edges based on individual reliability gains results in low-quality solution.

\begin{algorithm}[tb!]
	\caption{\small Hill Climbing}
	\label{algo:hc}
	\begin{algorithmic}[1]
		\REQUIRE source node $s$, target node $t$ in uncertain graph $\mathcal{G}=(V,E,p)$,
		a budget $k$ for new edges, a probability threshold $\zeta$.
		\ENSURE A set of $k$ edges $E_1$ (each with probability $\zeta$) to add in $\mathcal{G}$ for maximizing the $s$-$t$ reliability
		\STATE Construct a set of candidate edges $E^+=V\times V\setminus E$, each with probability $\zeta$
		\STATE $E_1 \gets \emptyset$
		\WHILE {$|E_1|<k$}
		\STATE  $e^* = \argmax_{e \in E^+\setminus E_1}\left[R\left(s,t,\left(V,E\cup E_1\cup\{e\},p\right)\right) \right.$  \\
		$\qquad \qquad \qquad \qquad \qquad \left. - R\left(s,t,\left(V,E\cup E_1,p\right)\right)\right]$
		\STATE $E_1 \gets E_1\cup\{e^*\}$ 
		\ENDWHILE
		\RETURN $E_1$
	\end{algorithmic}
\end{algorithm}

\vspace{-3mm}
\subsection{Hill Climbing Method}
\label{sec:hc}
\vspace{-1mm}
A better-quality solution
would be the hill climbing algorithm (Algorithm~\ref{algo:hc}):
It greedily adds the edge that provides the maximum marginal gain to the $s$-$t$ reliability at the
current round, until total $k$ new edges have been selected. In particular,
consider that a set $E_1\subseteq V\times V\setminus E$ of
new edges have been already included, in the next iteration the hill climbing baseline selects
a new edge $e \in V\times V\setminus (E\cup E_1)$, with $p(e)=\zeta$, such that:
\vspace{-2mm}
\begin{align}
& e^* = \argmax_{e \in V\times V\setminus (E\cup E_1)}\left[R\left(s,t,\left(V,E\cup E_1\cup\{e\},p\right)\right) \right. \nonumber \\
& \qquad \qquad \qquad \qquad \left. - R\left(s,t,\left(V,E\cup E_1,p\right)\right)\right]
\label{eq:4}
\end{align}
Since Problem~\ref{prob:single-st} is neither submodular nor supermodular, this
approach does not provide approximation guarantees.

\vspace{-0.6mm}
\spara{Time complexity.} Assuming the number of missing edges to be $\bigO(n^2)$,
coupled with MC sampling, the time complexity of each iteration of
this approach is $\bigO(n^2Z(n + m))$.
For total $k$ iterations, overall complexity is $\bigO(n^2kZ(n + m))$.

\vspace{-0.6mm}
\spara{Shortcomings.} Hill climbing also suffers from efficiency and accuracy issues. {\bf (1)} This is more inefficient compared to individual top-$k$ baseline.
{\bf (2)} For accuracy, hill climbing still suffers from
the {\em cold start} problem: At initial rounds, there would be several new edges
with marginal reliability gain zero (or, quite small), resulting in random selections,
which in turn produces sub-optimal solutions at later stages.
\vspace{-3mm}
\subsection{Centrality-based Method}
\label{sec:stat}
\vspace{-1mm}
Another intuitive approach is to find highly central nodes
in the input graph, and connect them by new edges if they are
not already connected, until the budget $k$ on new edges is exhausted.
In particular, we consider (1) degree centrality,
that is, nodes having higher aggregated edge probabilities considering all
incoming and outgoing edges; (2) betweenness centrality, that is, nodes having larger number of shortest paths passing through. Such nodes are also known as the {\em hub nodes}:
Connecting these hub nodes help in reducing network distances (as well
as improving reliability over uncertain graphs).

\vspace{-0.6mm}
\spara{Time complexity.} For degree centrality, it requires going through all nodes and
checking their in/out going edges, which costs $\bigO(m+n)$ time. To calculate the betweenness centrality of all nodes, Brandes' algorithm \cite{Bran01} takes $\bigO(nm)$. Then, it ranks
the nodes based on their aggregated edge probabilities, which consumes $\bigO(n\log n)$
time. 

\vspace{-0.6mm}
\spara{Shortcomings.} Although
the method (in particular, degree centrality) is efficient, and in general
improves the $s$-$t$ reliability, it is not customized for a specific $s$-$t$ pair.
This often results in low-quality solution.

\begin{algorithm}[tb!]
	\caption{\small Eigenvalue-based Method}
	\label{algo:EO}
	\begin{algorithmic}[1]
		\REQUIRE The adjacency matrix $A$ of the input uncertain graph $\mathcal{G}=(V,E,p)$
		\ENSURE A set of $k$ edges $E_1$ (each with probability $\zeta$) to add in $\mathcal{G}$ for maximizing the largest eigenvalue of the input matrix $A$
		\STATE Compute the largest eigenvalue $\lambda$ of the input matrix $A$
		\STATE Compute the maximum in-degree $d_{in}$ and out-degree $d_{out}$ of the input graph $\mathcal{G}$
		\STATE Find the subset of nodes $I$ with top-$(k+d_{in})$ left eigen-scores and the subset of nodes $J$ with top-$(k+d_{out})$ right eigen-scores.
		\STATE Connect the nodes from $I$ to $J$ (if no such edges exists in $\mathcal{G}$), which results in a set of edges $E^+$
		\STATE Select the top-$k$ edges $E_1$ in $E^+$ with largest ${\bf u}(i){\bf v}(j)$, where {\bf u} and {\bf v} are the corresponding left and right eigenvectors with the leading eigenvalue $\lambda$ of the original adjacency matrix $A$
		\RETURN $E_1$
	\end{algorithmic}
\end{algorithm}

\vspace{-3mm}
\subsection{Eigenvalue-based Method}
\label{sec:eigen}
\vspace{-1mm}
Wang et al. \cite{WCWF03} studied the importance of the largest eigenvalue of graph topology in
the dissemination process over real networks. To model the virus propagation in a network,
they assumed a fixed infection rate $\beta$ for an infected node to pass the virus to its neighbor,
and another fixed curing rate $\delta$ for an infected node. Then, they
proved that if $\frac{\beta}{\delta}<\frac{1}{\lambda }$, where $\lambda$ is the largest eigenvalue of the adjacency
matrix of this network, the virus will die out in this network. Therefore, one can optimize the leading eigenvalue
to control the virus dissemination in a network, e.g., with smaller $\lambda$, smaller curing rate $\delta$ is required
for the same infecting rate $\beta$. 

\begin{table}[tb!]
	\centering
	\caption{Table of notations (used in \S 4, 5, 6).}
	\vspace{-2mm}
	\begin{tabular} {c||l}
		\hline
		{\bf Notation} & {\bf Description}  \\
		\hline \hline
		$Z$ & number of samples in Monte Carlo sampling \\  \hline
		$l$ &  number of most reliable paths \\ \hline
		\multirow{2}{*}{$MRP(s,t,\mathcal{G})$} & the path with maximum probability of existence \\ & between $s$ and $t$ in $\mathcal{G}$ \\ \hline
		$\mathcal{P}(s,t,\mathcal{G})$ &  set of all paths from $s$ to $t$ in $\mathcal{G}$ \\ \hline
		$C(s)$ &  top-$r$ nodes with highest reliability from $s$ \\ \hline
		$C(t)$ &  top-$r$ nodes with highest reliability to $t$ \\ \hline
		$r$ & number of relevant nodes from(to)  $s$($t$) \\ \hline
		\multirow{2}{*}{$\mathbb{F}$} &  aggregation function over reliability of all $s$-$t$ \\ & pairs \\ \hline
		\hline
	\end{tabular}
	\vspace{1mm}
	\label{tab:notation_2}
	\vspace{-2mm}
\end{table}

Recently, Chen et al. studied the problem of maximizing the largest eigenvalue of
a network by edge-addition \cite{CTPEFF16}. They proved that the eigenvalue gain of adding a set of $k$ new edges
$E_1$ can be approximated by $\sum_{e_x \in E_1} {\bf u}(i_x){\bf v}(j_x)$, where ${\bf u}$ and ${\bf v}$
are the corresponding left and right eigenvectors with the leading eigenvalue of the original adjacency
matrix ($i_x$ and $j_x$ are the two end points of the new edge $e_x$).
They also proved that each $e_x$ in optimal $E_1$ has left endpoint from the subset of $(k+d_{in})$ nodes with the
highest left eigen-score ${\bf u}(i_x)$, and right endpoint from the subset of $(k+d_{out})$ nodes with the highest right
eigen-score ${\bf v}(j_x)$, where $d_{in}$ and $d_{out}$ are the maximum in-degree
and out-degree in the original graph, respectively.
Therefore, one can find the optimal $k$ new edges to increase the eigenvalue of the input graph by the following steps (Algorithm~\ref{algo:EO}):
First, calculate the largest eigenvalue of the input graph, and the corresponding left and right eigenvectors.
Then, compute the maximum in-degree and out-degree of this graph, and find the subset of nodes $I$ with
top-$(k+d_{in})$ left eigen-scores and the subset of nodes $J$ with top-$(k+d_{out})$ right eigen-scores.
Finally, connect the nodes from $I$ to $J$ (if no such edge exists in the original graph), and select
the top-$k$ pairs with largest eigen-scores ${\bf u}(i_x){\bf v}(j_x)$.

\vspace{-0.6mm}
\spara{Time complexity.} The first step can be solved with power iteration method in $\bigO(n)$ time.
Finding maximum in/out degrees takes $\bigO(n+m)$ time, and finding subset $I$ and $J$ requires $\bigO(n(d_{in}+k))$
and $\bigO(n(d_{out}+k))$ times, respectively, which can be written as $\bigO(nt)$, $t=max(k,d_{in},d_{out})$.
The final step consumes $\bigO(kt^2)$ time. Therefore, the overall time complexity is $\bigO(m+nt+kt^2)$.

\vspace{-0.6mm}
\spara{Shortcomings.} {\bf (1)} 
This method is not customized for a specific $s$-$t$ pair,
and may report low-quality solutions. {\bf (2)} To the best of our knowledge, there is no equivalent transformation from
virus propagation threshold $\frac{\beta}{\delta}$ to the $s$-$t$ reliability. Therefore, maximizing 
virus propagation may not be equivalent to maximizing the $s$-$t$ reliability.

\vspace{-2mm}
\section{A Simplified Problem: Improve \\ The Most Reliable Path}
\label{sec:relpath}
\vspace{-1mm}

\begin{algorithm}[tb!]
	\caption{\small Improve the Most Reliable Path}
	\label{algo:MRP}
	\begin{algorithmic}[1]
		\REQUIRE source node $s$, target node $t$ in uncertain graph $\mathcal{G}=(V,E,p)$,
		a budget $k$ for new edges, a probability threshold $\zeta$.
		\ENSURE A set of $k$ edges $E_1$ (each with probability $\zeta$) to add in $\mathcal{G}$ for maximizing the probability of the most reliable path from $s$ to $t$
		\STATE Color all existing edges in $\mathcal{G}$ as blue
		\STATE Add all missing edges to the graph, each with edge probablity $\zeta$, and color them as red. The new graph is $\overline{\mathcal{G}}$
		\STATE Convert $\overline{\mathcal{G}}$ into a weighted graph $G_0$ by assigning weight $w(e)=-\log p(e)$
		\STATE Make $k$ identical copies of $G_0$: $\{G_0, G_1, ..., G_k\}$
		\FOR {$j$ from $k$ to $0$}
		\STATE Remove all red edges from $G_k$
		\FOR {Every $0\leq i \leq k-1$}
		\FOR {Every red edge $e_j=(v_a,v_b)$ in $G_i$}
		\STATE Remove $e_j$ from $G_i$
		\STATE Draw a new edge from $v_a$ in $G_i$ to $v_b$ in $G_{i+1}$
		\ENDFOR
		\ENDFOR
		\ENDFOR
		\STATE Find the shortest paths $\{P_0, P_2,...,P_k\}$ from $s$ in $G_0$ to every $t$ in $G_i$($0\leq i\leq k$)
		\STATE $P =\argmin_{1\leq i \leq k}W(P_i)$
		\RETURN The set of red edges on $P$ as $E_1$
	\end{algorithmic}
\end{algorithm}

Due to limitations of baseline approaches as discussed in \S~\ref{sec:algo},
we now explore an orthogonal direction following the notion of the {\em most reliable path}.
The idea that we shall develop in this section will be the basis of our ultimate
solution (to be introduced in \S~\ref{sec:proposed}) for the budgeted reliability maximization problem.
A path between a source $s$ and a target node $t$ in an uncertain graph $\mathcal{G}$
is called the most reliable path $MRP(s,t,\mathcal{G})$ if the probability of that path
(i.e., product of edge probabilities on that path) is maximum
in comparison with all other paths between these two nodes.
\vspace{-3mm}
\begin{align}
& MRP(s,t,\mathcal{G}) = \argmax_{P \in \mathcal{P}(s,t,\mathcal{G})} \prod_{e\in P} p(e)
\vspace{-1mm}
\end{align}
$\mathcal{P}(s,t,\mathcal{G})$ denotes the set of all paths from $s$ to $t$ in $\mathcal{G}$.
The problem that we investigate here is a simplified version of our original problem
(i.e., Problem~\ref{prob:single-st}) as stated next.
\begin{problem}
[Single-source-target most reliable path improvement]
Given an uncertain graph $\mathcal{G}=(V,E,p)$, a source $s\in V$, a target $t\in V$, a probability threshold $\zeta\in(0,1]$,
and a small positive integer $k$, find the top-$k$ edges to add in $\mathcal{G}$, each new edge $e$ having probability $p(e)=\zeta$,
such that the probability of the most reliable path from $s$ to $t$ in the updated graph is maximized.
\vspace{-2mm}
\begin{align}
& E^*= \argmax_{E_1\subseteq V\times V\setminus E } \prod_{e\in MRP\left(s,t,\left(V,E\cup E_1,p\right)\right)} p(e) \nonumber  \\
&\text{s. t.} \qquad |E_1|=k; \qquad \text{and} \quad p(e)=\zeta \quad \forall e \in E_1
\vspace{-3mm}
\end{align}
\vspace{-5mm}
\label{prob:reliable-st}
\end{problem}

\begin{table*}[tb!]
	\vspace{-1mm}
	\scriptsize
	\parbox{.47\linewidth}{
		\centering
		\vspace{-1mm}
		\caption{Reliability gain and running time comparison without search space elimination. $k=10$, $\zeta=0.5$, {\em lastFM}}
		\label{tab:Comp_mrp}
		\vspace{-3mm}
		\begin{tabular} {l||c|l}
			\hline
			{\textsf Method}& {\textsf Reliability Gain} & {\textsf Running Time (sec)} \\
			\hline \hline
			Individual Top-$k$ & 0.27 & 39184  \\ \hline
			{\bf Hill Climbing} & {\bf 0.32} & {\bf 406512} \\ \hline
			Centrality-based  & 0.03 & 19 \\ 
			(degree) & & \\ \hline
			Centrality-based  & 0.11 & 2998 \\ 
			(betweenness) & & \\ \hline
			Eigenvalue-based & 0.09 & 213 \\ \hline
			{\bf Most Reliable Path} & {\bf 0.26} & {\bf 467} \\	\hline
			Individual Path Inclusion & 0.29 & 332\\
			(proposed method)         &      &   \\ \hline
			Batch-edge Selection & 0.31& 421 \\
			(proposed method)    &            &   \\
			\hline
		\end{tabular}
	}
	\hfill
	\parbox{.47\linewidth}{
		\centering
		\vspace{-1mm}
		\caption{Reliability gain and running time comparison after search space elimination. $k=10$, $\zeta=0.5$, $l=30$, $r=100$, {\em lastFM}.
			Time for search space elimination: 16 sec.}
		\label{tab:Comp}
		\vspace{-3mm}
		\begin{tabular} {l||c|l}
			\hline
			{\textsf Method}& {\textsf Reliability Gain} & {\textsf Running Time (sec)} \\
			\hline \hline
			Individual Top-$k$ & 0.27 & 136  \\ \hline
			Hill Climbing & 0.31 & 1256 \\ \hline
			Centrality-based  & 0.13 & 5 \\ 
			(degree) & & \\ \hline
			Centrality-based  & 0.21 & 21 \\ 
			(betweenness) & & \\ \hline
			Eigenvalue-based & 0.20 & 19 \\ \hline
			Most Reliable Path & 0.25 & 20 \\ \hline
			{\bf Individual Path Inclusion} & {\bf 0.30} & {\bf 16}\\
			(proposed method)         &      &   \\ \hline
			{\bf Batch-edge Selection} & {\bf 0.33}& {\bf 22} \\
			(proposed method)    &            &   \\
			\hline
		\end{tabular}
	}
	\vspace{-3mm}
\end{table*}

\begin{figure*}[htb]
	\centering
	\vspace{-0.5mm}
	\subfigure[\small {Input graph}]{
	\includegraphics[scale=0.5]{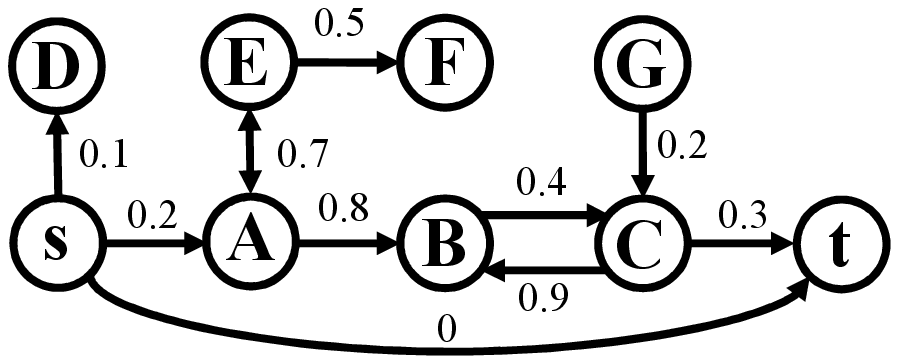}
	\label{fig:ograph}}
	\subfigure[\small {Adding relevant candidate edges}]{
	\includegraphics[scale=0.52]{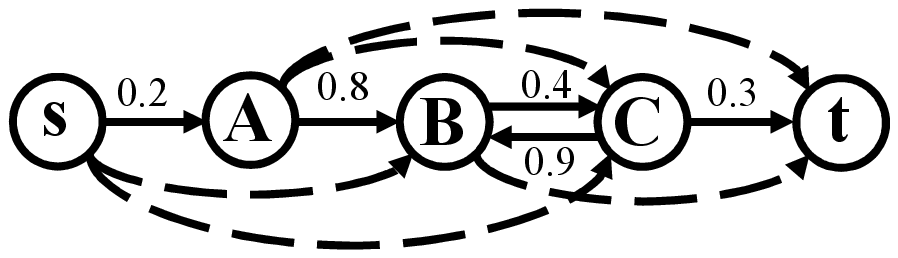}
	\label{fig:rgraph}}
	\subfigure[\small {Select top-$l$ reliable paths}]{
	\includegraphics[scale=0.54]{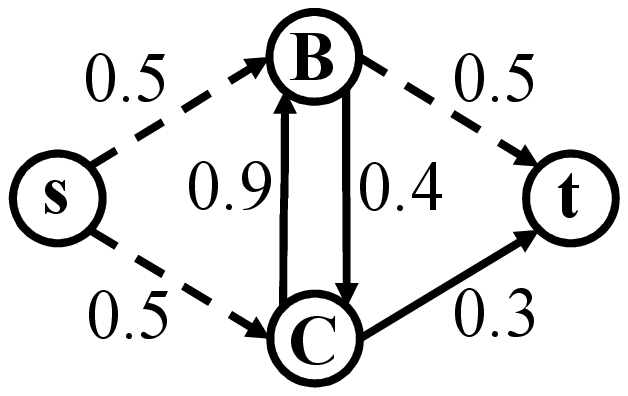}
	\label{fig:sgraph}}
	\vspace{-4mm}
	\caption{Run-through example for the proposed algorithm}
	\label{fig:example_single}
	\vspace{-6mm}
\end{figure*}

\vspace{-1.5mm}
Notice that the probability of the most reliable path from $s$ to $t$ cannot be larger than the $s$-$t$ reliability. Thus,
Problem~\ref{prob:reliable-st} might be considered as a {\em simplified} version of the budgeted reliability maximization problem.
Nevertheless, as reported in earlier studies \cite{KZK16,CWW10}, the most reliable path often provides a good approximation
to the reliability between a pair of nodes. Thus, our intuition is simple: If the solution of Problem~\ref{prob:reliable-st}
is more efficient and results in higher-quality top-$k$ edges (compared to baselines for the original budgeted reliability
maximization problem), then we can augment this idea (e.g., instead of the most reliable path, one may consider multiple
highly-reliable paths from $s$ to $t$) to develop even better-quality solution for the budged reliability maximization problem.

Fortunately, Problem~\ref{prob:reliable-st} can be solved exactly in polynomial time. We shall provide a constructive proof,
which can also be used as an algorithm for Problem~\ref{prob:reliable-st}.

\vspace{-0.5mm}
\begin{theor}
	Problem~\ref{prob:reliable-st} can be solved in polynomial time.
\end{theor}
\vspace{-2mm}
\begin{IEEEproof}
First, we color all existing edges in the input graph $\mathcal{G}$ as blue (Algorithm~\ref{algo:MRP}).
Then, we add all missing edges to the graph, each with edge probability $\zeta$
(thus, resulting in a complete graph), and color new edges as red. Name this new graph as $\overline{\mathcal{G}}$.
The goal of Problem~\ref{prob:reliable-st} is to find the most reliable path from $s$ to $t$
containing at most $k$ red edges (with zero or more blue edges), if any. Notice that we can convert the uncertain graph $\overline{\mathcal{G}}$
into an weighted graph $G_0$, which has same set of edges and nodes as $\overline{\mathcal{G}}$, and the weight of each edge $e$ in
$G_0$ is: $w(e)=-\log p(e)$.
Equivalently, we aim at finding the shortest path from $s$ to $t$ in $G_0$ containing at most $k$ red edges
(with zero or more blue edges), if any.

To find such paths, $k$ identical copies of $G_0$ are made (denoted as $G_0, G_1, G_2, \ldots, G_k$), and are updated as follows.
\vspace{-1.5mm}
\begin{enumerate}
\setlength\itemsep{0.01em}
\item	Remove all red edges from $G_k$.
\item	For every $0\leq i \leq k-1$ \\
		\blank{5mm} For every red edge $e_j = (v_a, v_b)$ in $G_i$ \\
		\blank{1cm} Remove $e_j$ from $G_i$ \\
        \blank{1cm} Draw a new edge from $v_a$ in $G_i$ to $v_b$ in $G_{i+1}$
\vspace{-5mm}
\end{enumerate}
Now, we employ the Dijkstra's algorithm to find the shortest paths from $s$ in $G_0$ to every $t$ in $G_i$ ($0\leq i \leq k$).
Each shortest path from $s$ in $G_0$ to $t$ in $G_i$ corresponds to a path in the original graph $\overline{\mathcal{G}}$
with at most $i$ red edges
. We refer to these paths (if they exist) as $P_0, P_1, \ldots, P_k$, respectively.

Consider a function $W$ that gets as input a path, and returns the aggregate weight of edges on that path.
If for every $1 \leq i \leq k$, we have $W(P_0) \leq W(P_i)$, then adding no $k' \leq k$ edges to $\mathcal{G}$
can improve the probability of the most reliable path from $s$ to $t$. Otherwise, we find $P =\argmin_{1\leq i \leq k}W(P_i)$, and consider all red edges in $P$.
Adding these edges to $\mathcal{G}$ will result in the maximum probability of the most reliable path from $s$ to $t$.

The time required for the above method is due to running the Dijkstra's algorithm for $k+1$ times over a graph with $(k+1)n$ nodes and $(k+1)n^2$ edges.
Hence, the overall time complexity of our method is $\bigO(k^2n^2 + k^2n\log(kn))$, which is polynomial in input size. The theorem follows.
\end{IEEEproof}

\vspace{-0.6mm}
\spara{Comparison with baselines.} As shown in Table \ref{tab:Comp_mrp}, solving the simplified most reliable path problem (Problem~\ref{prob:reliable-st})
is much faster than both baselines: Individual Top-$k$ and Hill Climbing for the original problem (Problem~\ref{prob:single-st}).
As expected, the improvement in $s$-$t$ reliability via most reliable path-based solution is 0.26, which is lower but comparable to that of Hill Climbing: 0.32.
However, the most reliable path approach terminates in 467 seconds, while Hill Climbing consumes about 4.7 days.
For other two baselines, Centrality-based and Eigenvalue-based, the most reliable path method significantly outperforms them in reliability gain. 

\vspace{-2mm}
\section{Proposed Solution: Budgeted \\ Reliability Maximization}
\label{sec:proposed}
\vspace{-1mm}

In this section, we present our method ultimately designed
for an effective and efficient solution to the single-source-target
budgeted reliability maximization (Problem~\ref{prob:single-st}).
Due to the success of the most reliable path technique
as detailed in \S~\ref{sec:relpath}, our final solution is developed
based on a similar notion by employing multiple reliable paths, and further
improved in two ways: {\bf (1)} Reduction of search space
by identifying only the most relevant candidate edges for a given $s$-$t$ pair,
and {\bf (2)} improving solution quality by considering {\em multiple}
highly reliable paths from  $s$ to $t$. In the following, we discuss
various steps of our framework, and demonstrate accuracy
and efficiency improvements against previous baselines.

\vspace{-3mm}
\subsection{Search Space Elimination}
\label{sec:search}

\subsubsection{Reliability-based Search Space Elimination}
\vspace{-1mm}
\label{sec:r_search}
In a sparse input graph $\mathcal{G}$, one can have as many as $\bigO(n^2)$ candidate edges.
However, given a specific $s$-$t$ query, all candidate edges may not be equally relevant.
In particular, let us consider two nodes $u$ and $v$: Both have low reliability either from source $s$,
or to target $t$; then adding an edge between $u$ and $v$ will not improve the $s$-$t$ reliability significantly.
Therefore, we select ``relevant'' candidate edges as follows (Algorithm~\ref{algo:sse}).
{\bf (1)} We find the top-$r$ nodes with the highest reliability from $s$. Similarly, we compute the top-$r$ nodes
having the highest reliability to $t$. Let us refer to these sets as $C(s)$ and $C(t)$, respectively.
Notice that $s\in C(s)$ and $t\in C(t)$. 
{\bf (2)} For two distinct nodes $u,v$, such that $u\in C(s)$, $v\in C(t)$, and $u,v$ are not connected in the
input graph $\mathcal{G}$, then we consider the new edge $(u,v)$, with edge probability $p(u,v)=\zeta$, as a candidate edge. We denote by $E^+$ the set of relevant candidate edges. Thus, we reduce the number of candidate edges from $\bigO(n^2)$ to only $\bigO(r^2)$.

The time complexity of this step is $\bigO(Z(n+m)+ n\log r + r^2)$. The first term is due to MC sampling to compute the
reliability of all nodes from $s$ and to $t$, and the second term is due to sorting all nodes based on these reliability
values.

\begin{algorithm}[tb!]
	\caption{\small Search Space Elimination.}
	\label{algo:sse}
	\begin{algorithmic}[1]
		\REQUIRE source node $s$, target node $t$ in uncertain graph $\mathcal{G}=(V,E,p)$, a number threshold $r$
		\ENSURE  A set of edges $E^+$ as the search space
		\STATE Find the top-$r$ nodes $C(s)$ with highest reliability from the source node $s$
		\STATE Find the top-$r$ nodes $C(t)$ with highest reliability to the target node $t$
		\STATE $E^+ \gets \{(u,v)|u\neq v, u\in C(s), v\in C(t), (u,v)\notin E \}$
		\RETURN $E^+$
	\end{algorithmic}
\end{algorithm}

\vspace{-2mm}
\subsubsection{Top-$l$ Most-Reliable Paths Selection}
\vspace{-1mm}
Given the success of the most reliable path-based approach (\S~\ref{sec:relpath}), we further improve it with
multiple highly reliable paths. Recent research has shown that what really matters in computing the
reliability between two nodes is the set of highly reliable paths between them \cite{KBGN18,CWW10,KZK16}.

On adding the relevant candidate edges $E^+$,
we refer to the updated graph as $\mathcal{G}^+=(V,E\cup E^+,p)$. Next, we find the top-$l$ most reliable paths from $s$ to $t$ with the Eppstein's algorithm \cite{E98,KBGN18} within $\bigO(m + n \log n + l)$.
If a new edge does not appear in any of these top-$l$ paths, it can be removed from $E^+$. This further
reduces the search space.
\vspace{-1mm}
\begin{exam}
Let us demonstrate ``search space elimination'' with Figure \ref{fig:example_single}. Suppose we set $r=3$, $l=3$, and $\zeta=0.5$.
First, we select the top-$3$ nodes with highest reliability from source $s$. Clearly, $\{s,A,B\}$ will be selected. Similarly, $\{B,C,t\}$ are the top-$3$ nodes with highest reliability to target $t$. Node $D$, $E$, $F$, and $G$ will be eliminated, and we obtain a graph presented in Figure \ref{fig:rgraph}. Then, we select top-3 most reliable paths between $s$ and $t$ after adding all missing edges (dotted lines) with given probability $\zeta=0.5$ in Figure \ref{fig:rgraph}. They will be $\{sBt,sCBt,sCt\}$ (in decreasing order). Node $A$ does not appear in any of these paths, and will be eliminated. Finally, we have a simplified graph shown in Figure \ref{fig:sgraph}.
\label{ex:1}
\vspace{-1mm}
\end{exam}
\vspace{-5mm}
\spara{Benefits of search space elimination.} As shown in Table~\ref{tab:Comp}, our search space elimination methods can save
about 99\% of running time for the baselines: Individual Top-$k$ and Hill Climbing without accuracy loss. For Centrality-based and Eigenvalue-based
baselines, both efficiency and accuracy get improved, because these baselines are now applied over a smaller and more relevant (to a specific $s$-$t$ pair)
subgraph. After including the time cost for conducting search space elimination: 16 seconds, the overall running time for most reliable path method and our
proposed algorithms can be reduced by over 70\% without accuracy loss.

\vspace{-2mm}
\subsection{Top-$k$ Edges Selection}
\vspace{-1mm}
Our next objective is to find the top-$k$ edges from the reduced set $E^+$ of candidate edges, so to maximize the $s$-$t$ reliability.
We formulate the problem as follows.

\begin{problem}
[Budgeted Path Selection]
Given the set $\mathcal{P}$ of the top-$l$ most reliable paths from $s$ to $t$ in the updated graph $\mathcal{G}^+$,
find a path set $\mathcal{P}^* \subseteq \mathcal{P}$ such that:
\vspace{-2mm}
\begin{align}
& \mathcal{P}^* = \argmax_{\mathcal{P}_1\subseteq \mathcal{P}} R(s,t,\mathcal{P}_1) \nonumber \\
& \text{s. t.} \qquad |\{e:e\in E^+ \cap \mathcal{P}_1\}| \leq k
\vspace{-3mm}
\end{align}
\label{prob:bps}
\end{problem}
\vspace{-7mm}

For Problem~\ref{prob:bps}, $R(s,t,\mathcal{P}_1)$ denotes the $s$-$t$ reliability on the subgraph induced by the path set $\mathcal{P}_1$.
In other words, we find a path set $\mathcal{P}^*\subseteq \mathcal{P}$ that maximize the $s$-$t$ reliability,
while also satisfying the constraint on $k$, the number of newly-added edges.
Unfortunately, this problem is \NP-hard as well, which can be proved by a reduction from the MAX $k$-COVER. Since the
proof is analogous to the one in Theorem~\ref{th:np}, we omit this for brevity. Instead, we design two practical and
effective solutions as given below.

\begin{algorithm}[tb!]
	\caption{\small Individual Path-based Edge Selection}
	\label{algo:IP}
	\begin{algorithmic}[1]
		\REQUIRE source node $s$, target node $t$ in uncertain graph $\mathcal{G}=(V,E,p)$,
		a budget $k$ for new edges, a probability threshold $\zeta$, a candidate number threshold $r$, a most reliable path number threshold $l$
		\ENSURE  A set of $k$ edges $E_1$ (each with probability $\zeta$) to add in $\mathcal{G}$ for maximizing the $s$-$t$ reliability
		\STATE Invoke Algorithm 4 to obtain the candidate edge set $E^+$
		\STATE $\mathcal{G}^+=(V,E\cup E^+,p)$, each edge in $E^+$ is assigned a probability of $\zeta$
		\STATE Find the top-$l$ most reliable paths $\mathcal{P}$ from $s$ to $t$ in $\mathcal{G}^+$
		\STATE $E_1 \gets \emptyset$, $\mathcal{P}_1 \gets \emptyset$
		\STATE Move those the paths which do not contain any edge in $E^+$ from $\mathcal{P}$ into $\mathcal{P}_1$
		\WHILE {$|E_1|<k$}
		\STATE $P^*= \argmax_{P \in \mathcal{P} \setminus \mathcal{P}_1}R(s,t,\mathcal{P}_1\cup\{P\})$
		\STATE $\mathcal{P}_1 \gets \mathcal{P}_1 \cup \{P^*\}$
		\STATE Extract the set of edges $E_{P^*}$ on path $P^*$
		\STATE $E_1 \gets E_1\cup (E_{P^*}\cap E^+)$
		\FOR {$P$ in $\mathcal{P}$}
		\STATE Extract the set of edges $E_{P}$ on path $P$
		\IF {$|E_1 \cup (E_P\cap E^+)|>k$}
		\STATE Remove $P$ from $\mathcal{P}$
		\ENDIF
		\ENDFOR
		\ENDWHILE
		\RETURN $E_1$
	\end{algorithmic}
\end{algorithm}

\begin{table*}[tb!]
	\scriptsize
	\vspace{-2mm}
	\parbox{.36\linewidth}{
		\centering
		\caption{\small Running time comparison for reliability-based search space elimination, $r=100$. We also report the number of samples ($Z$) required by {\em MC} and {\em RSS}.}
		\label{tab:rel_eli}	
		\vspace{-1mm}
		\begin{tabular} {c||c|c||c|c}
			\hline
			\multirow{2}{*}{Dataset}& \multicolumn{2}{c||}{MC sampling} & \multicolumn{2}{c}{RSS sampling} \\ \cline{2-5}
			& Z & Time (sec) & Z & Time (sec)  \\
			\hline \hline
			{\em lastFM}& 1000 & 16 & 250 & 9 \\
			{\em AS\_Topology} & 500 & 166 & 250 & 12\\
			{\em DBLP} & 750 & 498 & 250 & 98 \\
			{\em Twitter}  & 1000 & 439 & 500 & 114\\
			\hline
		\end{tabular}
	}
	\hfill
	\parbox{.6\linewidth}{
		\centering
		\caption{\small Running time comparison for top-$k$ edges selection, $r=100$. We also report the number of samples ($Z$) required by {\em MC} and {\em RSS}.}
		\label{tab:selec}	
		\vspace{0.5mm}
		\begin{tabular} {c||c||c|c|c||c||c|c|c}
			\hline
			\multirow{2}{*}{Dataset}& \multicolumn{4}{c||}{MC sampling time (sec)} & \multicolumn{4}{c}{RSS sampling time (sec)} \\ \cline{2-9}
			& $Z$ & HC & MRP & Batch-edge & $Z$ & HC & MRP & Batch-edge \\
			\hline \hline
			{\em lastFM} & 500 & 1256 & 20 & 22 & 250 & 708 & 15 & 16 \\
			{\em AS\_Topology} & 500 & 1508 & 23& 29 & 250 & 758 & 19& 22\\
			{\em DBLP} & 500 & 1818 & 34 & 50 & 250 & 1007 & 27 & 31\\
			{\em Twitter} & 500 & 1677 & 38 & 44 & 250 & 939 & 26 & 27\\
			\hline
		\end{tabular}
	}
	\vspace{-3mm}
\end{table*}

\vspace{-2mm}
\subsubsection{Individual Path-based Edge Selection}
\label{sec:ind_path}
\vspace{-1mm}
The algorithm is shown as Algorithm~\ref{algo:IP}.
First, we combine all paths from $\mathcal{P}$ that do not have any candidate edges from $E^+$ (line 5).
We refer to these paths as $\mathcal{P}_1$, and the subgraph induced by $\mathcal{P}_1$ as $G^*$.
Then, in each successive round, we iteratively include a remaining path $P^*$ from $\mathcal{P}\setminus\mathcal{P}_1$
into $G^*$ which maximally increases the reliability (estimated via MC-sampling) from $s$ to $t$ in $G^*$ (line 7), while still maintaining the budget
$k$ on the number of included candidate edges in $G^*$ . It can be formulated as:
\vspace{-2mm}
\begin{align}
& P^*= \argmax_{P \in \mathcal{P} \setminus \mathcal{P}_1}R(s,t,\mathcal{P}_1\cup\{P\})
\vspace{-4mm}
\end{align}
We ensure that the number of included candidate edges from $E^+$ in $\mathcal{P}_1$ does not exceed $k$ during the process (line 11-16).
The included candidate edges in $G^*$ are reported as our solution.

Let us denote by $n'$ and $m'$ the number of nodes and edges, respectively, in the subgraph induced by the top-$l$ most-reliable path set $\mathcal{P}$,
and $T$ the number of MC samples required in each iteration. We need at most $k$ iterations, thus the overall time complexity is $\bigO(kZ|P|(n'+m'))$.
\begin{algorithm}[tb!]
	\caption{\small Path Batch Construction}
	\label{algo:pb_c}
	\begin{algorithmic}[1]
		\REQUIRE A set of most reliable paths $\mathcal{P}$, a set of candidate edges $E^+$
		\ENSURE  A set of path batches $\mathcal{PB}$
		\STATE $\mathcal{PB} \gets \emptyset$
		\FOR {Path $P\in \mathcal{P}$}
		\STATE Extract the set of edges $E_P$ on path $P$
		\STATE Compute the label $L=E_P\cap E^+$ for path $P$
		\IF {$PB_L \notin \mathcal{PB}$}
		\STATE $PB_L \gets \emptyset$
		\STATE $\mathcal{PB}\gets \mathcal{PB}\cup \{PB_L\}$
		\ENDIF
		\STATE $PB_L \gets PB_L\cup\{P\}$
		\ENDFOR
		\RETURN $\mathcal{PB}$
	\end{algorithmic}
\end{algorithm}

\vspace{-2mm}
\subsubsection{Path Batches-based Edge Selection}
\label{sec:path_batch}
\vspace{-1mm}
The effectiveness of individual path selection can be improved by considering the relationships between paths
in $\mathcal{P}$. The intuitions are: (1) different paths can share same set of candidate edges;
(2) the candidate edge set of a path can be a subset of that for another path; and
(3) different paths may have different number of candidate edges to be included.

Therefore, we design a path batch-based (instead of individual path-based) edge selection algorithm.
First, we go through all paths in $\mathcal{P}$. If two paths share same set of candidate edges,
they will be put into the same ``path batch'' (Algorithm~\ref{algo:pb_c}). Each path batch is labeled by its candidate edge set (line 4, Algorithm~\ref{algo:pb_c}),
and in our algorithm we include a path batch in every round. In general, it follows the same procedure of Algorithm~\ref{algo:IP}, but invokes Algorithm~\ref{algo:pb_c} after line 5. 
In all of follow-up steps of Algorithm 5, we shall select path batch $PB$ instead of path $P$.
When evaluating the marginal gain of a path batch,
all other path batches whose candidate edge set is a subset of it, shall also be included in $G^*$ in the current round.
The marginal gain of this batch is normalized by the size of its candidate edge set.
The detailed procedure is shown in the following example.
\begin{exam}
Consider Example \ref{ex:1} and Figure \ref{fig:example_single}, we are now selecting top-$2$ edges from 3
candidate edges $\{sB,sC,$ $Bt\}$. If selecting paths individually, path $sBt$ has the highest marginal
gain 0.25 and will be selected in the first round. As budget $k=2$ is exhausted, the solution set is $\{sB,Bt\}$.
However, path $sCt$ has reliability gain 0.15, and only adds 1 new edge. Its marginal gain per new edge is higher
than that of $sBt$. Further, by considering it in batch path selection manner, including path $sCBt$ will also activate
path $sCt$
The reliability gain of adding
them in batch is 0.3075,
and the marginal gain per new edge is 0.1538, which is the winner of this round, and we find the optimal solution $\{sC,Bt\}$
in this example. The reliability gains for the other 2 possible solution are 0.28 for $\{sB,Bt\}$, 0.18 for $\{sB,sC\}$.
This demonstrates the effectiveness of path batch selection procedure.
\vspace{-1mm}
\end{exam}

\vspace{-1mm}
\spara{Benefits of path batches-based edge selection.} As shown in Table \ref{tab:Comp}, path batch selection and
Hill Climbing have similar reliability gain, 
while path batch selection consumes significantly less running time. Comparing with individual path inclusion, path batch selection
has some improvement in reliability gain, with comparable running time.
\vspace{-4mm}
\subsection{Improvement via Advanced Sampling}
\vspace{-1mm}
\label{sec:ad_sample}
Recently, several advanced sampling methods have been proposed for estimating $s$-$t$ reliability, including
lazy propagation \cite{LFZT17}, recursive sampling \cite{JLDW11}, recursive stratified sampling (RSS) \cite{LiYMJ16},
and probabilistic tree \cite{MCS17}. While our problem and the proposed solution are orthogonal to
the specific sampling method used, its efficiency can further be improved by employing more sophisticated
sampling strategies \cite{KKL19}. In particular, instead of MC sampling, we shall consider RSS in the experiments, both for our proposed method and for
the baselines.
%

The recursive stratified sampling \cite{LiYMJ16} partitions the probability space $\Omega$ into $r+1$ non-overlapping
subspaces ($\Omega_0$,...,$\Omega_r$) via selecting $r$ edges. In stratum $i$, we set the status of edge $i$ to 1,
the status of those edges before it as 0, and all other edges as undetermined. The probability $\pi_i$ of stratum $i$
can be calculated as the product of the absent probability $1-p(e)$ of all edges with 0 status, multiplied by the
probability of edge $i$. The sample size of stratum $i$ is set as $Z_i=\pi_i\cdot Z$, where $Z$ is the total sample size.
The algorithm recursively computes the sample size to each stratum and simplify the graph. It applies Monte Carlo
sampling on the simplified graph when the sample size of a stratum is smaller than a given threshold. Reliability is
then calculated by finding the sum of the reliabilities in all subspaces. The time complexity of recursive stratified sampling \cite{LiYMJ16} is same as that of the MC sampling,
i.e., $\bigO(Z(m+n))$, while the variance of the estimator is significantly reduced. Therefore, if we require same variance, recursive stratified sampling runs faster due to its smaller sample size $Z$.

\vspace{-1mm}
\spara{Benefits of recursive stratified sampling.} We compute the variance of an estimator by repeating experiments with different
number of samples ($Z$), and we consider the ratio $\rho_Z=\frac{V_Z}{R_Z}$ to decide if the estimator has converged over a given dataset.
Here, $V_Z$ is the average variance of repeating 100 different $s$-$t$ queries for
100 times, and $R_Z$ is the mean of reliability of these queries. The ratio of variance
to mean, also known as the index of dispersion, is a normalized measure of the dispersion of a dataset .
If $\rho_Z<0.001$, we conclude that the estimator has converged.

Table \ref{tab:rel_eli} and \ref{tab:selec} report the number of samples
($Z$) required for convergence in each dataset, together with the running time comparison. Clearly, applying recursive stratified sampling (RSS)
significantly reduces the running time of sampling-based methods. For reliability-based search space elimination, the sampling is conducted on the original graphs
(which are large in size). RSS requires about half of the sample size, compared to that of MC sampling, and reduces the running time by 50\%-90\%.
For top-$k$ edges selection, the sampling is applied over a simplified (smaller) subgraph, however the benefit of RSS over MC sampling
can still be up to 40\%.

\vspace{-2mm}
\section{Multiple-Source-Target \\ Reliability Maximization}
\label{sec:multi}
\vspace{-1mm}
In practice, queries may consist of multiple source and/or target nodes, rather than a single $s$-$t$ pair.
For example, in targeted marketing  \cite{KZK16,KKC18} via social networks,
the campaigner wants to maximize the information diffusion from a group of early adopters 
to a set of target customers. For such real-world applications, we extend our problem to adapt to multiple source/target nodes.
In particular, we focus on maximizing an aggregate function (e.g., average, maximum, minimum) over reliability of
{\em all} $s$-$t$ pairs.
\vspace{-0.5mm}
\begin{problem}
[Multiple-source-target budgeted reliability maximization]
Given an uncertain graph $\mathcal{G}=(V,E,p)$, a set of source nodes $S \subset V$, a set of target nodes $T \subset V$, a probability threshold $\zeta\in(0,1]$,
and a small positive integer $k$, find the top-$k$ edges to add in $\mathcal{G}$, each new edge having probability $p(e)=\zeta$,
such that an aggregate function $\mathbb{F}$ over reliability of all $s$-$t$ pairs ($s\in S$, $t\in T$) is maximized.
\vspace{-2.5mm}
\begin{align}
	& \displaystyle E^*= \argmax_{E_1\subseteq V\times V\setminus E } \underset{s,t\in S\times T}{\mathbb{F}}\left(R \left(s,t,\left(V,E\cup E_1,p\right)\right)\right) \nonumber  \\
	&\text{s. t.} \qquad |E_1|=k; \qquad \text{and} \quad p(e)=\zeta \quad \forall e \in E_1
\vspace{-2mm}
\end{align}
\vspace{-2mm}
\label{prob:multi}
\end{problem}

\vspace{-5mm}
Due to {\bf NP}-hardness of Problem~\ref{prob:single-st}, its generalization, Problem~\ref{prob:multi} is also {\bf NP}-hard.
In the following sections, we consider three widely-used aggregate functions: average, minimum, maximum; and design efficient solutions. 

\vspace{-4mm}
\subsection{Maximizing the Average Reliability}
\label{sec:max_avg}
\vspace{-1mm}
Our objective is:
\vspace{-3mm}
\begin{align}
\vspace{-5mm}
\displaystyle \argmax_{E_1\subseteq V\times V\setminus E } \frac{1}{|S||T|}\underset{s,t\in S\times T}{\sum}R\left(s,t,\left(V,E\cup E_1,p\right)\right)
\label{eq:max_avg}
\end{align}
Note that this is equivalent to maximizing the sum of reliability of all $s$-$t$ pairs.
For targeted marketing in social networks, a campaigner would like to
maximize the spread of information to the entire target group; and therefore, she would prefer
to maximize the average reliability.

Similar to the single-source-target budgeted reliability maximization problem, we first compute
the reliable sets from source and target nodes, that is, $C(s)$ for all $s \in S$, and $C(t)$
for all $t \in T$. Next, for each pair of distinct nodes $u,v$, such that $u\in C(s), \forall s \in S$,
$v\in C(t), \forall t \in T$, and $u,v$ are not connected in the input graph $\mathcal{G}$,
we consider a new edge $(u,v)$, having edge probability $p(u,v)=\zeta$, as a
relevant candidate edge. We denote by $E^+$ the set of relevant candidate edges,
and after adding them to $\mathcal{G}$, we refer to the updated graph as $\mathcal{G}^+=(V,E\cup E^+,p)$.

Now, for each $s$-$t$ pair, we identify the top-$l$ most reliable paths in $\mathcal{G}^+$.
Then we have total $|S||T|l$ paths in this set, and the path set might contain more than
$k$ new edges. Therefore, we employ the path batches-based edge selection method (\S~\ref{sec:path_batch}):
The algorithm iteratively includes path batches that maximize the marginal gain considering our current objective
function (Equation~\ref{eq:max_avg}), while maintaining the budget $k$ on the number of newly inserted
edges.

\vspace{-0.8mm}
\spara{Time complexity.} Let $\bigO(P_1)$ denote the time complexity of reliability-based search space elimination, $\bigO(P_2)$ denote that of top-$l$ most-reliable paths selection, and $\bigO(P_3)$ denote that of path batches-based edge selection, for the single-source-target case. The time complexity of the proposed algorithm for average multiple-source-target budgeted reliability maximization problem is $\bigO((|S|+|T|)P_1+|S||T|(P_2+P_3))$. We need to evaluate all nodes' reliability from/to each source/target, which results in the first term. The second term is due to applying top-$l$ path selection algorithm $|S||T|$ times for each $s$-$t$ pair, and the path set size will be $|S||T|$ times of that for single-source-target problem.
%

\vspace{-4mm}
\subsection{Maximizing the Minimum Reliability}
\label{sec:max_min}
\vspace{-1mm}
Our objective is:
\vspace{-3.5mm}
\begin{align}
\vspace{-5mm}
\displaystyle \argmax_{E_1\subseteq V\times V\setminus E } \underset{s,t\in S\times T}{\min}R\left(s,t,\left(V,E\cup E_1,p\right)\right)
\label{eq:max_min}
\end{align}
In other words, we aim at including $k$ new edges such that the reliability of the $s$-$t$ pair having the lowest reliability
(after the addition of $k$ edges) is maximized. In the targeted marketing setting, this can happen during complementary influence
maximization \cite{LCL15}, where multiple products are being campaigned simultaneously, and they are complementary in nature:
Buying a product could boost the probability of buying another. Now, consider that each source node (e.g., an early adopter)
is campaigning a different, but complementary product. The campaigner would prefer to maximize the
minimum spread of her campaign from any of the early adopters to any of her target
users, because only a small percentage of the users who have heard about a campaign will buy the corresponding product.

To solve this problem, we first estimate the $s$-$t$ reliability for each pair in $S\times T$ over the input graph $\mathcal{G}$.
We sort these $s$-$t$ pairs in ascending order in a priority queue based on their current reliability.
Next, in successive rounds, we keep improving the reliability of the pair having the smallest current reliability,
until the budget on $k$ new edges can be exhausted.
In particular, at any point in our algorithm, we know which $s$-$t$ pair has the minimum reliability.
We extract this pair from the top of the priority queue, and improve its reliability with the addition of
a batch of suitable, new edges. For this purpose, we employ our algorithm for the single-source-target pair
(discussed in \S~\ref{sec:proposed}). The batch size can be set as $k_1<<k$.
Note that the addition of new edges not only updates the reliability of the current
pair, instead this will also increase the reliability of other $s$-$t$ pairs.

Thus, after adding a batch of $k_1$ new edges, we re-compute the reliability of all $s$-$t$ pairs
and re-organize them in the priority queue. Once again, we extract the pair from the top of the priority queue,
and improve its reliability with the addition of $k_1$ suitable, new edges. We repeat the above steps.
Ultimately, we terminate the algorithm when we exhaust our budget of adding total $k$ new edges.

\vspace{-0.8mm}
\spara{Time complexity.} The time complexity of our algorithm is $\bigO((|S|+|T|)P_1+\frac{k}{k_1}AP_1+\frac{k}{k_1}P_2+P_3)$, $A=min(|S|,|T|)$. Similar to maximizing the average reliability, we need evaluating all nodes' reliability from/to each source/target. All $s$-$t$ pair's original reliability can also be known through this process. However, after improving the reliability of the currently selected $s$-$t$ pair by adding $k_1$ edges, we need to update the reliability of all $s$-$t$ pairs. This will happen $\frac{k}{k_1}$ times, and the cost is $\frac{k}{k_1}AP_1$, $A=min(|S|,|T|)$. The top-$l$ paths selection will be operated $\frac{k}{k_1}$ times. The complexity of executing top-$k_1$ edge selection is $\bigO(\frac{k_1}{k}P_3)$, and it will happen $\frac{k}{k_1}$ times. Thus, the total time cost for edge selection remains $\bigO(P_3)$.

\vspace{-4mm}
\subsection{Maximizing the Maximum Reliability}
\label{sec:max_max}
\vspace{-1mm}
Our objective function is:
\vspace{-3mm}
\begin{align}
\vspace{-2mm}
\displaystyle \argmax_{E_1\subseteq V\times V\setminus E } \underset{s,t\in S\times T}{\max}R\left(s,t,\left(V,E\cup E_1,p\right)\right)
\label{eq:max_max}
\vspace{-4mm}
\end{align}
In the targeted marketing scenario, let us again consider complementary influence
maximization \cite{LCL15}, where each source user (e.g., an early adopter)
is campaigning a different, but complementary product. However, each target
user is now a celebrity in Twitter. Hence, the campaigner wants at least
one target user to be influenced by one of her products. In other words,
the campaigner would be willing to maximize the spread of information from at least
one early adopter to at least one target customer.

Note that if $S\cap T \ne \phi$, the problem is trivial,
as the maximum reliability is already one. Therefore, below
we consider the case when $S\cap T = \phi$. A straightforward solution to our problem would be to separately consider each $s$-$t$ pair from $S\times T$, improve its reliability
by adding $k$ new, suitable edges. Then, we pick the pair which achieves the maximum final reliability, and report those $k$ new edges that
were selected for this $s$-$t$ pair. However, the time complexity of this approach is $\bigO(|S||T|)$ times to that of a single $s$-$t$ pair.

Next, we develop a more efficient algorithm without significantly affecting the quality. Our approach is similar to that of maximizing the
minimum reliability (discussed in \S~\ref{sec:max_min}). In each round, we maximize the reliability (by adding $k_1 << k$ new edges)
of the pair having the current maximum reliability. After this, we re-compute the reliability of all pairs, and again pick the one
which has the current maximum reliability. We terminate the algorithm when we exhaust our budget of adding total $k$ new edges.

\vspace{-0.8mm}
\spara{Time complexity.} The time complexity will be the same as that of maximizing the minimal reliability, which is $\bigO((|S|+|T|)P_1+\frac{k}{k_1}AP_1+\frac{k}{k_1}P_2+P_3)$, $A=min(|S|,|T|)$.

\vspace{-2mm}
\section{Related Work}
\vspace{-1mm}
%

\begin{table*} [tb!]
	\scriptsize
	\centering
	\vspace{-2.5mm}
	\caption{Properties of datasets, SPL denotes shortest path length, and C. Coe. denotes clustering coefficient}
	\vspace{-3mm}
	\begin{tabular} { l||c|c|c|c|c|c|c }
		\hline
		{\textbf{Dataset}}        & {\textbf{\#Nodes}}  & {\textbf{\#Edges}}  &  {\textbf{Edge Prob: Mean, SD, Quartiles}} & {\bf Type}  & {\bf Avg. SPL} & {\bf Longest SPL} & {\bf C. Coe.}\\ \hline \hline
		{\em Intel Lab Data}              & 54	            &  969	          & 0.33 $\pm$ 0.19, \{0.16, 0.27, 0.44\} & Device, Directed  & 2.0 & 7 & 0.71 \\
		{\em LastFM}              & 6\,899	            &  23\,696	          & 0.29 $\pm$ 0.25, \{0.13, 0.20, 0.33\} & Social, Undirected  & 7.1 & 24 & 0.13\\
		{\em AS\_Topology}	      & 45\,535	        &  172\,294	      & 0.23 $\pm$ 0.20, \{0.08, 0.21, 0.31\} & Device, Directed  & 3.2 & 9 & 0.36\\
		{\em DBLP}	      & 1\,291\,298	        &  7\,123\,632	      & 0.11 $\pm$ 0.09, \{0.05, 0.10, 0.14\} & Social, Undirected  & 6.7 & 20 & 0.63 \\
		{\em Twitter}	          & 6\,294\,565	        &  11\,063\,034	      & 0.14 $\pm$ 0.10, \{0.10, 0.10, 0.19\} & Social, Undirected  & 4.5 & 7 & 0.57\\ \hline
		{\em Random\_1} & 1\,000\,000 & 2\,500\,000 & 0.30 $\pm$ 0.04, \{0.15, 0.30, 0.45\} &  Synthetic, Undirected & 4.1 & 10 & 0.11 \\
		{\em Random\_2} & 1\,000\,000 & 5\,000\,000 & 0.30 $\pm$ 0.04, \{0.15, 0.30, 0.44\} &  Synthetic, Undirected & 4.0 & 9 & 0.12 \\
		{\em Regular\_1} & 1\,000\,000 & 2\,500\,000 & 0.30 $\pm$ 0.04, \{0.15, 0.30, 0.45\} &  Synthetic, Undirected & 11.1 & 22 & 0.56 \\
		{\em Regular\_2} & 1\,000\,000 & 5\,000\,000 & 0.30 $\pm$ 0.04, \{0.16, 0.31, 0.45\} &  Synthetic, Undirected & 10.8 & 21 & 0.56 \\
		{\em SmallWorld\_1} & 1\,000\,000 & 2\,500\,000 & 0.29 $\pm$ 0.04, \{0.14, 0.30, 0.44\} &  Synthetic, Undirected & 4.6 & 9 & 0.55 \\
		{\em SmallWorld\_2} & 1\,000\,000 & 5\,000\,000 & 0.30 $\pm$ 0.04, \{0.15, 0.30, 0.45\} &  Synthetic, Undirected & 4.5 & 9 & 0.59 \\
		{\em ScaleFree\_1} & 1\,000\,000 & 2\,500\,000 & 0.30 $\pm$ 0.04, \{0.15, 0.30, 0.45\} &  Synthetic, Undirected & 5.6 & 8 & 0.46 \\
		{\em ScaleFree\_2} & 1\,000\,000 & 5\,000\,000 & 0.29 $\pm$ 0.04, \{0.15, 0.29, 0.44\} &  Synthetic, Undirected & 4.8 & 9 & 0.48 \\ \hline
	\end{tabular}
	\vspace{-2mm}
	\label{tab:data}
\end{table*}

\vspace{-0.6mm}
\spara{Network design problems.} Network design, optimization, and modification are widely studied research topics,
where one modifies the network structure or attributes, targeting at some objective metrics or functions.

There exist many different metrics to characterize the ``goodness'' of the network, including
average shortest paths \cite{AJB00, BGLR05}, ratio of connected nodes \cite{SunS07},
relative size of the largest connected component and average size of other components \cite{AJB00},
network flow and delay \cite{MVRS18}, centrality \cite{PPT16}, average path length \cite{ManP15}, 
and spectral measures \cite{CTPEFF16}.
Spectral measures are derived from the adjacency and the Laplacian matrices of a graph.
For example, \cite{CTPEFF16} optimized the leading eigenvalue of a network by edge addition/deletion,
due to the finding that the leading eigenvalue of the underlying graph is the key metric in determining
the so-called ``epidemic threshold'' for a variety of dissemination models \cite{WCWF03}. However,
such global metric is not query-specific. In real-world, users may tend to optimize the network in a way
that is relevant {\em only} to themselves, e.g., a campaigner would like to improve the influence \cite{CDV19} of her product
to her target customers, but not that of all similar products (from other competitors), and neither to other users
who are not her targets. Moreover, many network metrics studied in the past cannot be easily
generalized to probabilistic scenarios (e.g., connected component size). Our objective, {\em reliability}, is a fundamental
metric to capture the probability that a given target node is reachable from a specific source node in an uncertain graph.
Furthermore, we show that it is possible to generalize our objective to multiple-source-target cases
in order characterize a larger region in the network.

The major network manipulation operations include node addition/deletion \cite{AJB00, BGLR05}, edge addition/deletion \cite{CTPEFF16, PPT16, CDV19},
edge rewiring \cite{BGLR05}, and updating edge weights \cite{MVRS18}. Our goal is to improve the reliability between $s$-$t$ pairs in a network:
In our application scenarios, adding new edges is usually more practical. For example, it is often not realistic to set up a new airport only to
improve the reliability of connections between two existing airports, rather establishing some new flights is much easier.
In this paper, we study the problem of maximizing the reliability between a given pair of nodes by adding a small number of new edges.
Altering the existing edge probabilities is not investigated here, and can be an interesting future research direction on this problem.

\vspace{-0.6mm}
\spara{Reliability in uncertain networks.}
Due to the {\bf \#P}-hardness of $s$-$t$ reliability estimation problem, various efficient sampling approaches have been proposed in the literature.
Monte Carlo (MC) sampling \cite{F86} is a fundamental approach, which samples $Z$ possible worlds from the input uncertain graph, and approximates
the $s$-$t$ reliability with the ratio of possible world in which $t$ is reachable from $s$.
One may combine MC sampling with BFS from the source node to further improve its efficiency \cite{JLDW11}.
\cite{LFZT17} proposed Lazy Propagation, which utilizes geometric distribution to avoid frequent probing of edges.
BFSSharing improves the efficiency with offline indexes. Recursive sampling \cite{JLDW11} and recursive stratified sampling \cite{LiYMJ16}
reduces the estimator variance by recursively partitioning the search space. Less samples are required for them to achieve the same variance as previous methods,
thereby improving the efficiency. More recently, ProbTree index \cite{MCS17} was designed to support faster $s$-$t$ reliability queries over uncertain graphs.
Our problem and the proposed solution are orthogonal to the specific sampling method used, we demonstrate in \S~\ref{sec:proposed} 
that its efficiency can be improved by employing
recursive stratified sampling.

Orthogonal directions to our problem include adaptive edge testing
\cite{FXPFW17} and crowdsourcing \cite{LPCX17} for reducing uncertainty. In this work, we focus on
improving the reliability of a $s$-$t$ pair by adding a limited number of new edges.

\vspace{-2mm}
\section{EXPERIMENTAL RESULTS}
\label{sec:exp}

\vspace{-1mm}
We perform experiments to demonstrate effectiveness, efficiency, scalability, and memory usage of our algorithms.
We report sensitivity analysis by varying all input parameters in this section.
The code is implemented in C++, executed on a single core, 40GB, 2.40GHz Xeon server.

\vspace{-2mm}
\subsection{Experimental Setup}
\vspace{-1mm}

\begin{table*}[tb!]
	\vspace{-1mm}
	\begin{center}
		\caption{Single-source-target reliability maximization on different real datasets. $k=10$, $\zeta=0.5$, $r=100$, $l=30$.}
		\vspace{-3mm}
		\begin{tabular} {c||c|c|c|c||c|c|c|c||c|c|c|c}
			\hline
			\multirow{2}{*}{\bf Dataset}& \multicolumn{4}{c||}{\bf Reliability Gain} & \multicolumn{4}{c||}{\bf Running Time (sec)} & \multicolumn{4}{c}{\bf Memory Usage (GB)} \\ \cline{2-13}
			& HC & MRP & IP & BE & HC & MRP & IP & BE & HC & MRP & IP & BE \\
			\hline \hline
			{\em lastFM}& 0.31 & 0.25 & 0.30 & {\bf 0.33} & 717 & 24 & {\bf 14} & 25 & 0.06 & {\bf 0.04} & {\bf 0.04} & {\bf 0.04} \\
			{\em AS\_Topology} & {\bf 0.42} & 0.40 & 0.41 & {\bf 0.42} & 785 & 30 & {\bf 26} & 32 & 0.30 & {\bf 0.28} & {\bf 0.28} & 0.29\\
			{\em DBLP} & {\bf 0.24} & 0.19 & 0.22 & {\bf 0.24} & 1105 & 125 & {\bf 118} & 129 & 6.9 & {\bf 6.2} & 6.5 & 6.5\\
			{\em Twitter}  & 0.13 & 0.11 & 0.15 & {\bf 0.19} & 1053 & 140 & {\bf 127} & 141 & 11.0 & {\bf 9.4} & 9.8 & 9.8\\
			\hline
		\end{tabular}
		\label{tab:vary_dataset}
	\end{center}
	\vspace{-3mm}
\end{table*}

\begin{table*}[tb!]
	\begin{center}
		\caption{Single-source-target reliability maximization on different synthetic datasets. $k=10$, $\zeta=0.5$, $r=100$, $l=30$.}
		\vspace{-3mm}
		\begin{tabular} {c||c|c|c|c||c|c|c|c||c|c|c|c}
			\hline
			\multirow{2}{*}{\bf Dataset}& \multicolumn{4}{c||}{\bf Reliability Gain} & \multicolumn{4}{c||}{\bf Running Time (sec)} & \multicolumn{4}{c}{\bf Memory Usage (GB)} \\ \cline{2-13}
			& HC & MRP & IP & BE & HC & MRP & IP & BE & HC & MRP & IP & BE \\
			\hline \hline
			{\em Random\_1}& {\bf 0.17} & 0.13 & 0.15 & {\bf 0.17} & 1142 & 142 & {\bf 120} & 145 & 5.8 & {\bf 5.2} & 5.3 & 5.4 \\
			{\em Random\_2}& {\bf 0.16} & 0.13 & {\bf 0.16} & {\bf 0.16} & 1240 & 160 & {\bf 131} & 171 & 7.1 & {\bf 6.7} & 6.9 & 6.9 \\
			{\em Regular\_1}& 0.23 & 0.15 & 0.16 & {\bf 0.24} & 609 & 54 & {\bf 38} & 69 & 5.1 & {\bf 4.7} & 4.8 & 4.9 \\
			{\em Regular\_2}& 0.21 & 0.16 & 0.16 & {\bf 0.22} & 697 & 84 & {\bf 54} & 91 & 6.8 & {\bf 6.4} & 6.6 & 6.6 \\
			{\em SmallWorld\_1}& {\bf 0.20} & 0.15 & 0.16 & 0.19 & 877 & 83 & {\bf 60} & 80 & 5.4 & {\bf 5.0} & 5.1 & 5.2 \\
			{\em SmallWorld\_2}& 0.18 & 0.14 & 0.17 & {\bf 0.19} & 967 & 104 & {\bf 79} & 101 & 7.0 & {\bf 6.4} & 6.5 & 6.5 \\
			{\em ScaleFree\_1}& 0.18 & 0.14 & 0.16 & {\bf 0.19} & 926 & 94 & {\bf 81} & 101 & 5.5 & {\bf 5.0} & 5.2 & 5.2 \\
			{\em ScaleFree\_2}& {\bf 0.17} & 0.14 & 0.15 & {\bf 0.17} & 972 & 116 & {\bf 93} & 120 & 6.9 & {\bf 6.4}  & 6.6 & 6.6 \\
			\hline
		\end{tabular}
		\label{tab:vary_syn_dataset}
	\end{center}
	\vspace{-7mm}
\end{table*}

\begin{table}[tb!]
	\centering
	\caption{Comparison with the exact solution (ES), $k=3$, $\zeta=0.33$, $r=54$, $l=30$, {\em Intel Lab Data}.}
	\vspace{-3mm}
	\begin{tabular} {c||c|c}
		\hline
		{\bf Method} & {\bf Reliability Gain} & {\bf Running Time (sec)} \\
		\hline \hline
		ES & 0.252& 19189\\
		IP & 0.222& 8\\
		{\bf BE} & {\bf 0.237}& {\bf 12}\\
		\hline
	\end{tabular}
	\vspace{1mm}
	\label{tab:es}
	\vspace{-5mm}
\end{table}

\spara{Real-world Datasets.} We use 5 real-world graphs, consisting of 3 social and 2 device networks (Table \ref{tab:data}). {\bf (1)}{\em  Intel Lab Data} (\url{http://db.csail.mit.edu/labdata/labdata.html}). It is a collection of sensor communication data with 54 sensors deployed in the Intel Berkeley Research Lab between February 28th and April 5th, 2004. {\bf (2)} {\em  LastFM} (\url{www.last.fm}). It is a musical social network, where users listen to musics, and share them with friends. An edge between two users exists if they communicate at least once. {\bf (3)} {\em AS\_Topology} (\url{http://data.caida.org/datasets/topology/ark/ipv4/}). An autonomous system (AS) is a collection of connected Internet Protocol (IP) routing prefixes under the control of one or more network operators on behalf of a single administrative entity, e.g., a university. The AS connections are established with BGP protocol. It may fail due to various reasons, e.g., failure when one AS updates its connection configuration to ensure stricter security setting, while some of its peers can no longer satisfy it, or some connections are cancelled manually by the AS administrator. We downloaded one network snapshot per month, from January 2008 to December 2017. {\bf (4)} {\em DBLP} (\url{https://dblp.uni-trier.de/xml/}). It is a well-known collaboration network. We downloaded it on March 31, 2017. Each node is an author and edges denote their co-author relations. {\bf (5)} {\em Twitter} (\url{http://snap.stanford.edu/data/}). This is a widely used social network: Nodes are users and edges are re-tweets.

\vspace{-0.8mm}
\spara{Synthetic Datasets.} In order to study the effects of network properties to the performance of our algorithms, we generate 8 synthetic datasets, with the help of NetworkX package (\url{https://networkx.github.io}). They can be categorized into following 4 kinds of networks, each having 2 instances with different number of edges. {\bf (1)} {\em Random}. We generate the random networks with the well-known Erd\H{o}s-R\'{e}nyi model, that is, the edge between every pair of nodes exists with a fixed proabablity $p$. Here, the $p$ value for {\em Random\_1} is $5\times10^{-6}$, and the one for {\em Random\_2} is $1\times10^{-5}$. The random network tends to have a small average shortest path length, and a small clustering coefficient. The degrees of nodes in general follow Poisson distribution. {\bf (2)} {\em Regular}. We adopt the $k$-regular networks model here. In a $k$-regular network, every node has same degree of $k$. In particular, the $k$ value for {\em Regular\_1} is 5, and the one for {\em Regular\_2} is 10. In a regualr network, the average shortest path length is usually high, and the clustering coefficient is also high. {\bf (3)} {\em SmallWorld}. In a small-world network, most nodes are not neighbors of one another, but the neighbors of any given node are likely to be neighbors of each other, and most nodes can be reached from every other node by a small number of hops or steps. Thus, it always has a small shortest path length, and a high clustering coefficient. The widely-used Watts-Strogatz model for generating small-world networks starts with a $k$-regular lattice, and re-writes its edge connections with probability $p$ to obtain a small-world graph. We adopt $k=5$ and $k=10$ for our two instances, respectively. The $p$ is set to $0.3$.
{\bf (4)} {\em ScaleFree}. A scale-free network is a network whose degree distribution follows a power law. It usually has relatively smaller shortest path length, and higher clustering coefficient. Our scale-free networks are generated with Barab\'{a}si-Albert preferential attachment model, where a graph of $n$ nodes is grown by attaching new nodes each with $m$ edges that are preferentially attached to existing nodes with high degree. For {\em ScaleFree\_2}, the $m$ is set to be $5$. In order to maintain a consistent number of edges as all previous networks, we make slight change in the source code to allow alternating $m=2$ and $m=3$ during the generation. Real-world social networks are mostly both small-world and scale-free.

\begin{table*}[tb!]
	\vspace{-1mm}
	\parbox{.48\linewidth}{
		\centering
		\caption{Reliability gain and running time comparison with varying budget on \#new edges $k$. $\zeta=0.5$, $r=100$, $l=30$, {\em LastFM}.}
		\vspace{-3mm}
		\begin{tabular} {c||c|c|c|c||c|c|c|c}
			\hline
			\multirow{2}{*}{$k$}& \multicolumn{4}{c||}{\bf Reliability Gain} & \multicolumn{4}{c}{\bf Running Time (sec)} \\ \cline{2-9}
			& HC & MRP & IP & BE & HC & MRP & IP & BE \\
			\hline \hline
			3 & 0.24 & {\bf 0.25} & 0.21 & {\bf 0.25} & 257 & 11 & {\bf 10} & 13\\
			5 & {\bf 0.27} & 0.25 & 0.26 & {\bf 0.27} & 386 & 17 & {\bf 12} & 17\\
			8 & {\bf 0.29} & 0.25 & 0.27 & 0.28 & 525 & 20 & {\bf 13} & 19\\
			10 & 0.31 & 0.25 & 0.30 & {\bf 0.33} & 717 & 24 & {\bf 14} & 25\\
			15 & 0.34 & 0.25 & 0.32 & {\bf 0.35} & 953 & 26 & {\bf 17} & 28\\
			20 & 0.35 & 0.25 & 0.35 & {\bf 0.37} & 1378& 30& {\bf 20} & 32 \\
			30 & 0.38 & 0.25 & {\bf 0.39} & {\bf 0.39} & 2010& 37& {\bf 26} & 36 \\
			50 & 0.40 & 0.25 & 0.40 & {\bf 0.41} & 4005& 48& {\bf 34} & 44 \\
			\hline
		\end{tabular}
		\vspace{1mm}
		\label{tab:k_last}
	}
	\hfill
	\parbox{.48\linewidth}{
		\centering
		\caption{Reliability gain and running time comparison with varying budget on \#new edges $k$. $\zeta=0.5$, $r=100$, $l=30$, {\em DBLP}.}
		\vspace{-3mm}
		\begin{tabular} {c||c|c|c|c||c|c|c|c}
			\hline
			\multirow{2}{*}{$k$}& \multicolumn{4}{c||}{\bf Reliability Gain} & \multicolumn{4}{c}{\bf Running Time (sec)} \\ \cline{2-9}
			& HC & MRP & IP & BE & HC & MRP & IP & BE \\
			\hline \hline
			3 & 0.19 & 0.19 & 0.19 & {\bf 0.20} & 373 & 96 & {\bf 95} & 100 \\
			5 & {\bf 0.21} & 0.19 & 0.20 & {\bf 0.21} & 576 & 103& {\bf 97} & 106\\
			8 & 0.22 & 0.19 & 0.21 & {\bf 0.23} & 923& 111 & {\bf 107} & 110\\
			10 & {\bf 0.24} & 0.19 & 0.22 & {\bf 0.24} & 1097 & 117 & {\bf 112} & 121\\
			15 & 0.24 & 0.19 & 0.23 & {\bf 0.25} & 1504 & 127 & {\bf 119} & 126\\
			20 & {\bf 0.26} & 0.19 & 0.25 & {\bf 0.26} &  1974 & 136 & {\bf 125} & 131\\
			30 & {\bf 0.26} & 0.19 & {\bf 0.26} & {\bf 0.26} &  3091& 148 & {\bf 131} & 136\\
			50 & 0.27 & 0.19 & {\bf 0.28} & {\bf 0.28} &  5102& 162 & {\bf 139} & 142\\
			\hline
		\end{tabular}
		\label{tab:k_dblp}
		\vspace{1mm}
	}
	\vspace{-2mm}
\end{table*}

\begin{table*}[tb!]
	\parbox{.48\linewidth}{
		\centering
		\caption{Reliability gain, running time comparison with varying probability $\zeta$ on new edges. $k=10$, $r=100$, $l=30$, {\em AS\_Topology}.}
		\vspace{-3mm}
		\begin{tabular} {c||c|c|c|c||c|c|c|c}
			\hline
			\multirow{2}{*}{$\zeta$}& \multicolumn{4}{c||}{\bf Reliability Gain} & \multicolumn{4}{c}{\bf Running Time (sec)} \\ \cline{2-9}
			& HC & MRP & IP & BE & HC & MRP & IP & BE \\
			\hline \hline
			0.3 & 0.26 & 0.23 & 0.25 & {\bf 0.27} & 780 & 28 & {\bf 25} & 30 \\
			0.4 & {\bf 0.34} & 0.31 & 0.32 & 0.33& 774 & {\bf 27} & 28 & 29\\
			0.5 & {\bf 0.42} & 0.40 & 0.41 & {\bf 0.42} & 785 & 30 & {\bf 26} & 32 \\
			0.6 & {\bf 0.51} & 0.48 & 0.50 & {\bf 0.51} &801 & {\bf 32} & {\bf 32} & 37 \\
			0.7 & 0.59 & 0.57 & 0.58 & {\bf 0.60} & 810 & 37 & {\bf 35} & 40 \\
			1 & {\bf 0.84} & 0.82 & {\bf 0.84} & {\bf 0.84} & 855 & 47 & {\bf 40} & 50 \\
			\hline
		\end{tabular}
		\vspace{1mm}
		\label{tab:zeta_bio}
	}
	\hfill
	\parbox{.48\linewidth}{
		\centering
		\caption{Reliability gain, running time comparison with varying probability $\zeta$ on new edges. $k=10$, $r=100$, $l=30$, {\em Twitter}.}
		\vspace{-3mm}
		\begin{tabular} {c||c|c|c|c||c|c|c|c}
			\hline
			\multirow{2}{*}{$\zeta$}& \multicolumn{4}{c||}{\bf Reliability Gain} & \multicolumn{4}{c}{\bf Running Time (sec)} \\ \cline{2-9}
			& HC & MRP & IP & BE & HC & MRP & IP & BE \\
			\hline \hline
			0.3 & 0.10 & 0.07 & 0.10 & {\bf 0.12}& 1003 & 139 & {\bf 134} & 137 \\
			0.4 & 0.12 & 0.09 & 0.12 & {\bf 0.15}& 1025& 138 & {\bf 134} & 140 \\
			0.5 & 0.13 & 0.11 & 0.15 & {\bf 0.19} & 1053 & 140 & {\bf 136} & 141 \\
			0.6 & 0.17 & 0.15 & 0.19 & {\bf 0.24} & 1136 & 142 & {\bf 137} & 142 \\
			0.7 & 0.22& 0.17 & 0.27 & {\bf 0.29} & 1175 & 143 & {\bf 137} & 143\\
			1& {\bf 0.52} & 0.50 & 0.51 & {\bf 0.52} & 1235 & 146 & {\bf 140} & 146 \\
			\hline
		\end{tabular}
		\label{tab:zeta_twitter}
		\vspace{1mm}
	}
	\vspace{-2mm}
\end{table*}

\begin{table*}[tb!]
	\begin{center}
		\caption{Analysis with different probabilities on new edges. $k=10$, $r=100$, $l=30$, {\em Twitter}.}
		\vspace{-3mm}
		\begin{tabular} {c||c|c|c|c||c|c|c|c||c|c|c|c}
			\hline
			{\bf New edge}& \multicolumn{4}{c||}{\bf Reliability Gain} & \multicolumn{4}{c||}{\bf Running Time (sec)} & \multicolumn{4}{c}{\bf Memory (GB)} \\ \cline{2-13}
			{\bf probabilities}& HC & MRP & IP & BE & HC & MRP & IP & BE & HC & MRP & IP & BE \\
			\hline \hline
			rand$(0,1)$ & 0.17 & 0.13 & 0.16 & {\bf 0.18} & 1049 & 139 & {\bf 134} & 141 & 11.0 & {\bf 9.4} & 9.8 & 9.8\\
			rand$(0.2,0.6)$ & 0.13	& 0.10 & 0.11 & {\bf 0.14} & 1019 & 132 & {\bf 127} & 134 & 10.9 & {\bf 9.4} & 9.8 & 9.8\\
			rand$(0.4,0.8)$ & {\bf 0.21} & 0.17 & 0.20 & {\bf 0.21} & 1052 & 140 & {\bf 134} & 143 & 11.0 & {\bf 9.4} & 9.8 & 9.8 \\
			$N(0.5,0.038)$ & 0.19 & 0.14 & 0.16 & {\bf 0.20} & 1036 & 135 & {\bf 133} & 136 & 11.0 & {\bf 9.4} & 9.8 & 9.8\\
			\hline
		\end{tabular}
	\end{center}
	\label{tab:vary_zeta}
	\vspace{-4mm}
\end{table*}

\vspace{-0.8mm}
\spara{Edge probability models.} Our problems and solutions are {\em orthogonal to the specific way of assigning edge probabilities}. We adopt some widely-used models for generating edge probabilities in our evaluation. {\bf (1)} {\em Intel Lab Data} and {\bf (3)} {\em AS\_Topology}. The edge probabilities in these two datasets are real probabilities. For {\em Intel Lab Data}, the probabilities on edges denote the percentages of messages from a sender successfully reached to a receiver.
For {\em AS\_Topology}, once an AS connection (i.e., an edge) is observed for the first time, we calculate the ratio of snapshots containing this connection within all follow-up snapshots as the probability of the existence for this edge. {\bf (2)} {\em LastFM}. The probability on any edge is the inverse of the out-degree of the node from which that edge is outgoing. {\bf (4)} {\em DBLP} and {\bf (5)} {\em Twitter}. We assign the edge probability following $1-e^{-t/\mu}$, which is an exponential cdf of mean $\mu$ to a count $t$ \cite{JLDW11}. In {\em DBLP}, $t$ denotes the count of the collaborations between two authors. In {\em Twitter}, $t$ is the count of re-tweet actions. We set $\mu=20$. {\bf (6) Synthetic Datasets.} For 4 kinds of synthetic datasets, we assign a probability for each edge uniformly at random from range $(0,0.6]$.

\vspace{-0.8mm}
\spara{Queries.} For each dataset and single-source-target queries, we select 100 different $s$-$t$ pairs.
In practice, if two nodes are too close to each other, their original reliability will be naturally high;
thus, it might be unnecessary to improve their reliability further.
Thus, we select a source node uniformly at random, and find all its neighbors within 3-5 hops.
A target node is chosen from those neighbors randomly.

For multiple-source-target queries, we first generate a single-source-target query $s$-$t$. Then, for that $s$, we find all its neighbors that are within 5-hops away, and randomly select $q$ of them into the source set $S$ (i.e., source set size=$q$). Similarly, we pick $q$ of the within 5-hop neighbors of $t$ as the target set $T$, uniformly at random. We ensure that the source set and the target set do not overlap. Finally, 100 different source-target sets are generated.

\vspace{-0.8mm}
\spara{Parameters setup.} {\bf (1)} {\sf Budget on \#new edges ($k$)}. For single $s$-$t$ queries, we vary $k$ from 5 to 50, and use 10 as default.
For multiple-source-target queries, we vary $k$ from 10 to 500, use 100 as default. {\bf (2)} {\sf Probability on new edges ($\zeta$)}. We vary
$\zeta$ from 0.3 to 0.7, and use 0.5 as default. {\bf (3)} {\sf Number of candidate nodes ($r$)}. We vary $r$ from 20 to 300, and use 100 as default.
{\bf (4)} {\sf Number of most-reliable paths ($l$)}. We vary $l$ from 10 to 50, and use 30 as default. {\bf (5)} {\sf \#Sources and \#targets}.
For multiple-source-target queries, we vary source and target set sizes from 3 to 500.
{\bf (6)} {\sf The ratio of $\frac{k_1}{k}$}.
For multiple-source-target case with Max and Min aggregate functions, we further have a parameter $k_1$ as the budget for the current selected pair.
We vary $k_1$ from 5\% to 30\% of $k$, and use 10\% as default. {\bf (7)} {\sf Distance constraint for new edges}. In practice,
some missing edges cannot be candidate edges due to physical constraints. For example, in our case study in Introduction,
only short distance connections ($\leq 15$ meters) are allowed to be established.
In social networks, if two users have no common friends, it is unlikely that they will start communicating.
In current experiments, we assume that a missing edge can be added only if its two endpoints are $\leq h$-hops in the input graph. We vary $h$ from 2 to 5, and use 3 as default.
For real-world applications, one can easily set this constraint based on her requirements.
%


\vspace{-0.8mm}
\spara{Competing methods.} For single-source-target query, our ultimate method: path batches-based edge selection
(BE) is compared with individual path-based edge selection (IP), most reliable path (MRP), and our best baseline: Hill Climbing (HC). For baselines, the reliability gain is our major concern. Although HC is not efficient, it outperforms others in reliability gain (Tables~\ref{tab:Comp_mrp} and \ref{tab:Comp}).

For multiple-source-target case, we employ Hill Climbing (HC) , Eigenvalue-based Optimization (EO) \cite{CTPEFF16}, and two more recent methods, ESSSP \cite{PPT16} and IMA \cite{CDV19}, as competitors. Both ESSSP and IMA follow the same manner of adding a budget of new edges into the graph, each with a  fixed probability. The former aims at reducing the sum of expected shortest path length of each source-target pair, while the later attempts to increase the influence spread of the source nodes in the target nodes.

Moreover, on the smallest dataset, {\em Intel Lab Data}, we have the exact solution (ES) as a competitor, which enumerates all possible combinations of $k$ missing edges, and find the cone with highest reliability gain.
All of them are coupled with our search space elimination strategy (\S~\ref{sec:r_search}) and an
advanced sampling method: RSS (\S~\ref{sec:ad_sample}).

\vspace{-0.8mm}
\spara{Performance Metrics.} {\bf (1)} {\sf Reliability gain}.
We compute reliability gain due to $k$ new edges for each pair of source and target nodes, and report the average reliability gain over 100 distinct $s$-$t$ pairs.
{\bf (2)} {\sf Running time}. We report the end-to-end running time, averaged over 100 queries.
{\bf (3)} {\sf Memory usage}. We report the average memory usage of running each query.

\vspace{-2mm}
\subsection{Single-source-target results}
\label{sec:single_exp}

\vspace{-1mm}
\spara{Comparison with the exact solution.} The exact solution (ES) enumerates all possible combinations of $k$ missing edges, and finds the one with the highest reliability gain. The number of missing edges can reach $\bigO(n^2)$ in sparse graph, and results in $\binom{n^2}{k}$ possible choices, which makes it extremely inefficient in larger graphs. However, due to the small size of {\em Intel Lab Data}, we can apply such exhaustive search, and empirically compare with our proposed solution, BE. We follow the setting used in our case study in \S~\ref{sec:case}, that is, only 3 new short distance ($\leq 15$ meters) links are allowed to be established, each have the average probability 0.33.
30 distinct pairs of sensors, which are remote and with lower original reliabilities, are selected as queries.

As shown in Table \ref{tab:es}, our proposed solution, BE, exhibits very close performance against the exact solution (ES) in reliability gain. It returns same set of edges as ES, in 25 out of 30 queries. However, the running time of BE are at least three orders of magnitude faster than ES. This demonstrates both the effectiveness and the efficiency of our methods.

\begin{table*}[tb!]
	\scriptsize
	\parbox{.48\linewidth}{
		\centering
		\caption{\small Reliability gain and running time comparison with varying \#candidate nodes $r$. $k=10$, $\zeta=0.5$, $l=30$, {\em lastFM}. Time 1 denotes the time cost for search space elimination, and Time 2 is the time cost for top-$k$ edges selection.}
		\vspace{-3mm}
		\begin{tabular} {c||c|c|c|c||c||c|c|c|c}
			\hline
			\multirow{2}{*}{$r$}& \multicolumn{4}{c||}{\bf Reliability Gain} & {\bf Time 1} &  \multicolumn{4}{c}{\bf Time 2 (sec)} \\ \cline{2-5} \cline{7-10}
			& HC & MRP & IP & BE & {\bf (sec)} & HC & MRP & IP & BE \\
			\hline \hline
			20 & {\bf 0.29} & 0.25 & 0.27 & {\bf 0.29} & 6 & 223 & 5 & {\bf 2}& 9\\
			50 & 0.30 & 0.25 & 0.29 & {\bf 0.32} & 8 & 399 & 8 &  {\bf 3} & 10\\
			80 & 0.31 & 0.25 & 0.30 & {\bf 0.33} & 8 & 587 & 11 &  {\bf 4}& 13\\
			100& 0.31 & 0.25 & 0.30 & {\bf 0.33} & 9 & 708 & 15 & {\bf 5} & 16\\
			150 & 0.31 & 0.25 & 0.30 & {\bf 0.33}& 13 & 816 & 21 & {\bf 5} & 17\\
			200 & 0.31 & 0.25 & 0.30 & {\bf 0.33} & 19 & 849 & 26 & {\bf 7}& 17 \\
			300  & 0.31 & 0.25 & 0.30 & {\bf 0.33} & 29 & 898 & 35 & {\bf 9} & 19 \\
			\hline
		\end{tabular}
		\vspace{1mm}
		\label{tab:r_last}
	}
	\hfill
	\parbox{.48\linewidth}{
		\centering
		\caption{\small Reliability gain and running time comparison with varying \#candidate nodes $r$. $k=10$, $\zeta=0.5$, $l=30$, {\em DBLP}. Time 1 denotes the time cost for search space elimination, and Time 2 is the time cost for top-$k$ edges selection.}
		\vspace{-3mm}
		\begin{tabular} {c||c|c|c|c||c||c|c|c|c}
			\hline
			\multirow{2}{*}{$r$}& \multicolumn{4}{c||}{\bf Reliability Gain} & {\bf Time 1} &  \multicolumn{4}{c}{\bf Time 2 (sec)} \\ \cline{2-5} \cline{7-10}
			& HC & MRP & IP & BE & {\bf (sec)} & HC & MRP & IP & BE \\
			\hline \hline
			20 & 0.19 & 0.18 & 0.19  & {\bf 0.20} & 58 & 488 & {\bf 11} & 13 & 14\\
			50 & {\bf 0.22} & 0.19 & 0.20 & 0.21 & 71 & 650 & 19 & {\bf 14} & 20\\
			80 & 0.22 & 0.19 & 0.23 & {\bf 0.23} & 80 & 822 & 23 & {\bf 18} & 25 \\
			100 & {\bf 0.24} & 0.19 & 0.22 & {\bf 0.24} & 98 & 1007 & 27 & {\bf 22} & 31\\
			150 & {\bf 0.24} & 0.19 & 0.22 & {\bf 0.24} & 135 & 1339 & 37 & {\bf 25}& 35 \\
			200 & {\bf 0.24} & 0.19 & 0.22 & {\bf 0.24} & 190 & 1458 & 46 & {\bf 28}& 38\\
			300 & {\bf 0.24} & 0.19 & 0.22 & {\bf 0.24} & 411 & 1519 & 59 & {\bf 32 }& 43 \\
			\hline
		\end{tabular}
		\vspace{1mm}
		\label{tab:r_dblp}
	}
	\vspace{-2mm}
\end{table*}

\begin{table}[tb!]
	\vspace{-2.5mm}
	\begin{center}
		\caption{Reliability gain and running time comparison with varying distance $d$ between query nodes. $k=10$, $\zeta=0.5$, $r=100$, $l=30$, {\em AS\_Topology}.}
		\vspace{-3mm}
		\begin{tabular} {c||c|c||c|c}
			\hline
			\multirow{2}{*}{$d$}& \multicolumn{2}{c||}{\bf Reliability Gain} & \multicolumn{2}{c}{\bf Running Time (sec)}  \\ \cline{2-5}
			& {\quad HC } & {BE} & {\quad HC\quad} &{\em BE} \\
			\hline \hline
			2  & 0.29 & 0.31 & 669 & 140\\
			3  & 0.42 & {\bf 0.43} & 844 & 142\\
			4  & {\bf 0.39} & {\bf 0.39} & 830 & 133 \\
			5  & 0.21 & 0.24 & 471 & 128\\
			6  & 0.11 & 0.13 & 420 & 129\\
			\hline
		\end{tabular}
		\vspace{1mm}
		\label{tab:vary_d}
	\end{center}
	\vspace{-3mm}
\end{table}

\begin{table}[tb!]
	\begin{center}
		\caption{Reliability gain and running time comparison with varying distance constraint $h$ for new edges. $k=10$, $\zeta=0.5$, $r=100$, $l=30$, {\em Twitter}.}
		\vspace{-3mm}
		\begin{tabular} {c||c|c||c|c}
			\hline
			\multirow{2}{*}{$h$}& \multicolumn{2}{c||}{\bf Reliability Gain} & \multicolumn{2}{c}{\bf Running Time (sec)}  \\ \cline{2-5}
			& {\quad HC} & {\quad BE} & {\quad HC\quad} &{BE} \\
			\hline \hline
			2  & 0.11 & {\bf 0.14} & 661 & 130\\
			3  & 0.13 & {\bf 0.19} & 1053 & 141\\
			4  & 0.17 & {\bf 0.21} & 1615 & 166 \\
			5  & 0.19 & {\bf 0.22} & 1970 & 178\\
			\hline
		\end{tabular}
		\vspace{1mm}
		\label{tab:vary_h}
	\end{center}
	\vspace{-5mm}
\end{table}

\begin{table}[tb!]
	\small
	\centering
	\caption{\small Reliability gain and running time comparison with varying \#most-reliable paths $l$. $k=10$, $\zeta=0.5$, $r=100$, {\em Twitter}.}
	\vspace{0.5mm}
	\begin{tabular} {c||c|c||c|c}
		\hline
		\multirow{2}{*}{$\l$}& \multicolumn{2}{c||}{\bf Reliability Gain} & \multicolumn{2}{c}{\bf Running Time (sec)} \\ \cline{2-5}
		& IP & BE & IP & BE \\
		\hline \hline
		10 & 0.11 & {\bf 0.12} & {\bf 117} & 124 \\
		20 & 0.15 & {\bf 0.19} & {\bf 127} & 133\\
		30 & 0.15 & {\bf 0.19} &{\bf 136} & 141 \\
		40 & 0.15 & {\bf 0.19} & {\bf 148} & 150\\
		50 & 0.15 & {\bf 0.19} & {\bf 159} & 160 \\
		\hline
	\end{tabular}
	\label{tab:l_twitter}
	\vspace{-2mm}
\end{table}

\begin{table}[tb!]
	\centering
	\vspace{-1mm}
	\caption{Scalability analysis of BE. $k=10$, $\zeta=0.5$, $r=100$, $l=30$, {\em Twitter}.}
	\vspace{-3mm}
	\begin{tabular} {c||c|c|c}
		\hline
		\multirow{2}{*}{\bf \# Nodes}
		& \multirow{2}{*} {\bf Reliability Gain} & {\bf Running Time } & {\bf Memory Usage} \\
		& & {\bf (sec)} & {\bf (GB)} \\
		\hline \hline
		1M & 0.15 & 101 & 6.8\\
		2M & 0.17 & 109 & 5.7\\
		3M & 0.18 & 115 & 6.8\\
		4M & 0.19 & 122 & 7.9\\
		5M & 0.20 & 130 & 8.8\\
		6M & 0.19 & 141 & 9.8 \\
		\hline
	\end{tabular}
	\vspace{1mm}
	\label{tab:sca}
	\vspace{-6mm}
\end{table}

\vspace{-0.8mm}
\spara{Comparison of all competing methods on different real datasets with default parameters.} In Table \ref{tab:vary_dataset}, we present the reliability gain  obtained by four methods,
and the corresponding running time and memory usage, on various datasets with default parameters. Clearly, our ultimate method, path batches-based edge selection (BE)
outperforms others. For reliability gain, it wins on all datasets. On {\em Twitter}, the advantage of BE is more prominent.
The reason is that {\em Twitter} is a sparser graph compared to other datasets,
and the highly reliable paths connecting source to target are more likely to contain more than one missing edges ---
this fact enhances the impact of path batches. Individual path-based edge selection (IP)
always has lower reliability gain compared to BE. The polynomial-time solution, MRP for the restricted version of our problem has the lowest
reliability gain among these methods, as expected. 

Considering the running time, IP is the best one.
However, BE is only about 10-20 seconds slower than IP across all the datasets.
Both of them are about an order of magnitude faster than the baseline HC.
The memory usages of IP and BE are similar, while MRP costs slightly less memory.

\vspace{-0.8mm}
\spara{Comparison of all competing methods on different synthetic datasets with default parameters.} Similar to previous part, we present the results on synthetic datasets in Table \ref{tab:vary_syn_dataset}. Our ultimate method, BE, still outperforms others on all datasets. For reliability gain, it first confirms our finding on the real datasets that the reliability gain tends to be higher on sparser graphs, regardless of the kind of network. Further, it can be observed that we can achieve higher reliability gain on regular networks. It is well-known that establishing a few short cut edges can sharply reduce the average shortest path length in a network, and transforming it gradually into a small-world graph. The original path length is higher in a regular graph, which allows more for improvement.

The running time on random graphs is the highest, while that of regular graphs is lowest. The top-$r$ candidate nodes, $C(s)$ and $C(t)$, tend to be farther from $s$ or to $t$ on regular graphs. Since our $s$-$t$ query pairs are 3-5 hops away, $C(s)$ and $C(t)$ will have more overlap on regular graphs, which can reduce the number of candidate edges. Furthermore, clustering coefficients are high on regular graphs, thus more edges have existed from $C(s)$ to $C(t)$. This again reduces the number of candidate edges. These are the reasons for the smaller running time of our methods on regular graphs. The random graphs have the contrary properties, therefore it is slower to excute our algorithms there.

%


\begin{table*}[tb!]
	\vspace{-1mm}
	\begin{center}
		\caption{Reliability gain and running time comparison for multiple-source-target pairs. $k=100$, $\zeta=0.5$, $r=100$, $l=30$, $\frac{k_1}{k}=10\%$, {\em Twitter} ({\em Min.}).}
		\vspace{-3mm}
		\begin{tabular} {c||c|c|c|c|c||c|c|c|c|c}
			\hline
			\multirow{2}{*}{\bf \#Source:\#Target}&  \multicolumn{5}{c||}{\bf Reliability Gain} & \multicolumn{5}{c}{\bf Running Time (sec)}  \\ \cline{2-11}
			& {HC} & {EO} & ESSSP & IMA & {BE} & {HC} & {EO} & ESSSP & IMA & {BE} \\
			\hline \hline
			3:3 & 0.34 & 0.29 &0.35& 0.26 & {\bf 0.38} & 1662 & 384 & 726&658 & {\bf 358}\\
			10:10 & 0.27 & 0.19 & 0.33& 0.20&{\bf 0.36} & 2140 & {\bf 979} &1289&1058 & 1007  \\
			50:50 & 0.19 & 0.12 & 0.24&0.19&{\bf 0.28} & 4051 & 2910 &2946& {\bf 2310}  & 3049  \\
			100:100 & 0.15 & 0.06& 0.18& 0.11& {\bf 0.19} & 7144 & 5766 & 6690& 6302 & {\bf 5708} \\
			200:200 & 0.12 & 0.02 & 0.13&0.05&{\bf 0.16} & 13615 & {\bf 8711} & 9899& 8898& 8981\\
			500:500 & 0.12 & 0.04 & 0.12&0.05&{\bf 0.15} & 30111 & {\bf 17198} &22166&19454& 18082 \\
			\hline
		\end{tabular}
		\vspace{1mm}
		\label{tab:multi_min}
	\end{center}
	\vspace{-4mm}
\end{table*}
\begin{table*}[tb!]
	\vspace{-1mm}
	\begin{center}
		\caption{Reliability gain and running time comparison for multiple-source-target pairs. $k=100$, $\zeta=0.5$, $r=100$, $l=30$, $\frac{k_1}{k}=10\%$, {\em Twitter} ({\em Max.}).}
		\vspace{-3mm}
		\begin{tabular} {c||c|c|c|c|c||c|c|c|c|c}
			\hline
			\multirow{2}{*}{\bf \#Source:\#Target}&  \multicolumn{5}{c||}{\bf Reliability Gain} & \multicolumn{5}{c}{\bf Running Time (sec)}  \\ \cline{2-11}
			& {HC} & {EO} & ESSSP & IMA & {BE} & {HC} & {EO} & ESSSP & IMA & {BE} \\
			\hline \hline
			3:3 & {\bf 0.27} & 0.20 & 0.19 & 0.21 & {\bf 0.27} & 1682 & 404 & 772 & 662 & {\bf 377}\\
			10:10 & 0.24 & 0.17 & 0.16& 0.21 &{\bf 0.25} & 2381 & {\bf 982} & 1222 & 1091 &1071 \\
			50:50 & 0.24 & 0.16 & 0.17 & 0.20&  {\bf 0.28} & 4051 & 2955 &3042 & {\bf 2510}& 3366  \\
			100:100 & 0.23 & 0.16& 0.14& 0.19 & {\bf 0.26} & 7144 & {\bf 5822} & 6315 & 6335 & 6101  \\
			200:200 & 0.19 & 0.14 &0.15 &0.19  & {\bf 0.23} & 13615 & {\bf 8801} & 9538 & 9044 & 9114\\
			500:500 & 0.20 & 0.12 & 0.11& 0.18 & {\bf 0.22} & 30111 & {\bf 17699} & 20994 & 19049 & 18988\\
			\hline
		\end{tabular}
		\vspace{1mm}
		\label{tab:multi_max}
	\end{center}
	\vspace{-4mm}
\end{table*}
\begin{table*}[tb!]
	\vspace{-1mm}
	\begin{center}
		\caption{Reliability gain and running time comparison for multiple-source-target pairs. $k=100$, $\zeta=0.5$, $r=100$, $l=30$, $\frac{k_1}{k}=10\%$, {\em Twitter} ({\em Avg.}).}
		\vspace{-3mm}
		\begin{tabular} {c||c|c|c|c|c||c|c|c|c|c}
			\hline
			\multirow{2}{*}{\bf \#Source:\#Target}&  \multicolumn{5}{c||}{\bf Reliability Gain} & \multicolumn{5}{c}{\bf Running Time (sec)}  \\ \cline{2-11}
			& {HC} & {EO} & ESSSP & IMA & {BE} & {HC} & {EO} & ESSSP & IMA & {BE} \\
			\hline \hline
			1:1 &  0.22 &  0.23 &  0.21 &  {\bf 0.25} &  {\bf 0.25} &  2003 &  246  &  398 &  334 &  {\bf 182} \\
			3:3 & 0.21 & {\bf 0.24} & 0.23& {\bf 0.24}& {\bf 0.24} & 4221 & 369 & 701 & 641 & {\bf 239}\\
			10:10 & 0.17 & 0.19 & 0.17&0.20& {\bf 0.21} & 28948 & 944 & 1196 & 1115 & {\bf 787} \\
			50:50 & 0.14 & {\bf 0.17} & 0.15 & {\bf 0.17} & {\bf 0.17} & 131487 &  2856 & 2899 & 2401 &{\bf 2321}  \\
			100:100 & 0.12 & 0.13& 0.12 & 0.14 & {\bf 0.15} & 194449 & {\bf 2978} & 6449 & 6100 & 4662  \\
			200:200 & - & 0.09 & 0.09 & 0.11 & {\bf 0.12} & - &   8795 & 9647 & 8810  &  {\bf 7812}\\
			500:500 & - & 0.06 & 0.06 & 0.08 &{\bf 0.10} & - &  17189 & 20154 & 18994 & {\bf 13908}\\
			\hline
		\end{tabular}
		\label{tab:multi_avg}
	\end{center}
	\vspace{-5mm}
\end{table*}

\vspace{-0.8mm}
\spara{Varying the budget $k$ on \#new edges.} We present the results on {\em LastFM} and {\em DBLP} datasets in Tables \ref{tab:k_last} and \ref{tab:k_dblp}, respectively.
The reliability gain tends to increase with larger $k$.
Such growth is more significant when $k$ is small, for example the reliability gain increases from 0.27 to 0.33 when $k$ increases from 5 to 10, while only 0.02 increase
can be obtained when permitting $k$ from 20 to 30, on {\em LastFM}. The reliability gain nearly saturates at $k$=20 on {\em DBLP}.
On both datasets, BE outperforms others in reliability gain, no matter how large is $k$. The reliability gain of MRP converges at the beginning, because we only consider the most reliable path in this restricted version.
A path containing larger number of new edges tends to have longer length and lower probability, thus it is unlikely to be the most reliable path.

For running time,
MRP, IP, and BE are comparable, and all of them can finish within 200 seconds with the largest $k=50$. HC is $\approx$100$\times$ slower. The running time of MRP increases faster than IP and BE with larger $k$, since it requires $k$ copies of the original graph to find the most reliable paths with exactly 0 to $k$ missing edges,
although this does not help improve the solution quality in practice. 

\vspace{-0.8mm}
\spara{Varying probability $\zeta$ on new edges.} The experimental results on {\em AS\_Topology} and {\em Twitter} are provided in Tables \ref{tab:zeta_bio} and \ref{tab:zeta_twitter}, respectively.
The reliability grows almost linearly with the probability threshold $\zeta$. Sometimes, the growth rate may be even higher (e.g., on {\em Twitter}).
The reason is that the optimal solution set of edges may change with different $\zeta$ (Observation \ref{ob:zeta}), and a sharp increase
may happen when shifting from a set of edges to another (Example \ref{ex:char}). The running times of all the methods are not sensitive to different $\zeta$.
However, with larger $\zeta$, the running time slightly increases.

Table 16 provides additional analysis about the probabilities on new edges. Here, instead of a fixed threshold $\zeta$, we allow different probabilities on different new edges. Probabilities on new edges are generated uniformly at random in different range, or generated following normal distribution $N(0.5,0.038)$ (99\% of value generated are in range $(0,1)$). It can be viewed that the results are very similar to all our previous study with fixed threshold $\zeta$. {\em This confirms that our proposed algorithm, BE works well even when
different probabilities for the missing edges are provided as input.}

\vspace{-0.9mm}
\spara{Varying \#candidtae nodes $(r)$.}
In reliability-based search space elimination, we only keep the
top-$r$ nodes $C(s)$ with the highest reliability from $s$, and the top-$r$ nodes $C(t)$
with the highest reliability to $t$. $C(s)$ and $C(t)$ are candidate node sets, and only those missing edges from a node in $C(s)$ to a node in $C(t)$
will be considered as candidate edges. As demonstrated in Table \ref{tab:r_dblp}, small $r$ incurs low-quality result,
due to the excessive elimination. The accuracy does not keep improving if $r$ exceeds 80 and 100, respectively for {\em LastFM} and {\em DBLP}.
We find out that $r=100$ is sufficient for all the methods to work on all datasets in our experiments.

{\sf Time 1} denotes the time cost for search space elimination, and {\sf Time 2} is the time cost for top-$k$ edges selection. As shown in Tables \ref{tab:r_last} and \ref{tab:r_dblp}, when varying $r$, {\sf Time 1} 
increases sharply with larger $r$. Although the time cost of checking all nodes' reliability from/to a node
is not relevant to $r$, we need to add at most $\bigO(r^2)$ missing edges after determining the candidate nodes.
Since we shall also verify the distance constraint before adding a missing edge,
the time cost of adding edges is non-trivial. When $r\leq100$, the increasing rate for the running time of search space
elimination is modest. Together with the previous finding that $r=100$ can ensure a good accuracy,
we set $r=100$ as default in other experiments.

The running times for top-$k$ edge selection ({\sf Time 2})
for methods IP and BE increase little with larger $r$, since they estimate the reliability gain of missing edges only on the subgraph
induced by a few most-reliable paths. {\sf Time 2} for MRP, on the other hand, increases linearly
with $r$, since the size of each copy of graph is linear to $r$. The time cost of edge selection by HC
is also linear to $r$ at the beginning, and slows down with larger $r$. This is because that, although
the time complexity of sampling is linear to the graph size theoretically, when coupling with BFS search,
low-reliability nodes added later are less frequent to be explored during the sampling.

\vspace{-0.8mm}
\spara{Varying distance $(d)$ between query nodes $s$ and $t$.}
We further select queries where each $s$-$t$ pair are exactly $d$-hops
away in the input graph. As shown in Table \ref{tab:vary_d}, the original reliability decreases with larger $d$.
And the reliability gain at $d=3$ and $d=4$ is about the highest, for both HC and BE.

The running time is small either with too large or too small $d$. For small $d$, the candidate node sets $C(s)$ and $C(t)$ are likely to have a large overlap, thus less missing
edges are found. However, the distance between nodes in $C(s)$ and $C(t)$ tends to increase with larger $d$,
thus the distance constraint may forbid many missing edges from being added into the graphs. The running time of HC
is more sensitive to $d$, since it iterates over each new edge.



\vspace{-0.8mm}
\spara{Varying distance constraint $(h)$ for newly added edges.}
We constrain that a missing edge can only be added if the distance between its two endpoints
in the original graph is at most $h$ hops. Smaller $h$ prevents more edges from being added.
As shown in Table \ref{tab:vary_h}, with larger $h$,
we can obtain more reliability improvement. However, this allows many remote links
to be established, which may not be realistic in practice.
Moreover, many candidate edges also increase the running time, both for HC and BE.

\vspace{-0.9mm}
\spara{Varying \#most reliable paths $(l)$.} Table \ref{tab:l_twitter} demonstrate the sensitivity
analysis of our IP and BE methods to the number of most reliable paths, $l$.
The reliability gain increases with larger $l$, and saturates at around $l=30$. The running time is linear to $l$.
Thus, we set $l=30$ as default in the rest of our experiments.

\vspace{-0.8mm}
\spara{Scalability analysis.} We conduct scalability analysis of our method, BE
by varying the graph size on the largest dataset, {\em Twitter}. We select 1M, 2M, 3M, 4M, 5M, and 6M nodes
uniformly at random to generate 6 subgraphs, and apply our algorithm on them. Table \ref{tab:sca} shows that
the running time and the memory usage are both linear to the graph size, which confirms good scalability of BE.
\vspace{-2mm}
\subsection{Multiple-source-target results}
\vspace{-1mm}
\label{sec:multi_exp}
\vspace{-1.5mm}

\spara{Varying \#source-target nodes.} We present the results on the largest dataset, {\em Twitter}, in Tables \ref{tab:multi_min}, \ref{tab:multi_max}, and \ref{tab:multi_avg}, for the aggregate functions: {\em Minimum}, {\em Maximum}, and {\em Average}, respectively.  The baselines ESSSP and IMA can also run in single-source-target scenario, and we keep their results in the first row of Table~\ref{tab:multi_avg}. Our purposed method, BE, significantly outperforms the baselines, HC, EO, ESSSP, and IMA  in reliability gain, and runs at least 40$\times$ and 2$\times$ faster than HC, with {\em Average} and {\em Minimum/Maximum} aggregate functions, respectively.
The running times of EO, ESSSP, IMA, and BE are comparable.  In general, the running time of our method, BE is almost linear to the number
of source/target nodes. 

Furthermore, our method results in higher reliability gain with {\em all} 3 aggregate functions when comparing with EO, especially for
{\em Minimum} and {\em Maximum}. This is because EO is not query-specific. EO optimizes the leading
eigenvalue of a graph, which is a global metric and may have
little to do with the query pair having {\em Minimum} or {\em Maximum} reliability. The performance of IMA algorithm is closed to our method with {\em Average} aggregate function, since its objective, influence spread, can be considered as a variant of average aggregated reliability (see \S~\ref{sec:m_app}). When we have only one source-target pair, the objective of IMA becomes exactly the same as ours, which results in the same reliability gain, as shown in Table~\ref{tab:multi_avg}. The performance of ESSSP is always worse than our method.

\begin{figure}[t!]
	\centering
	\vspace{-2mm}
	\subfigure[\scriptsize {Reliability gain, {\em Twitter}}]
	{\includegraphics[scale=0.146,angle=270]{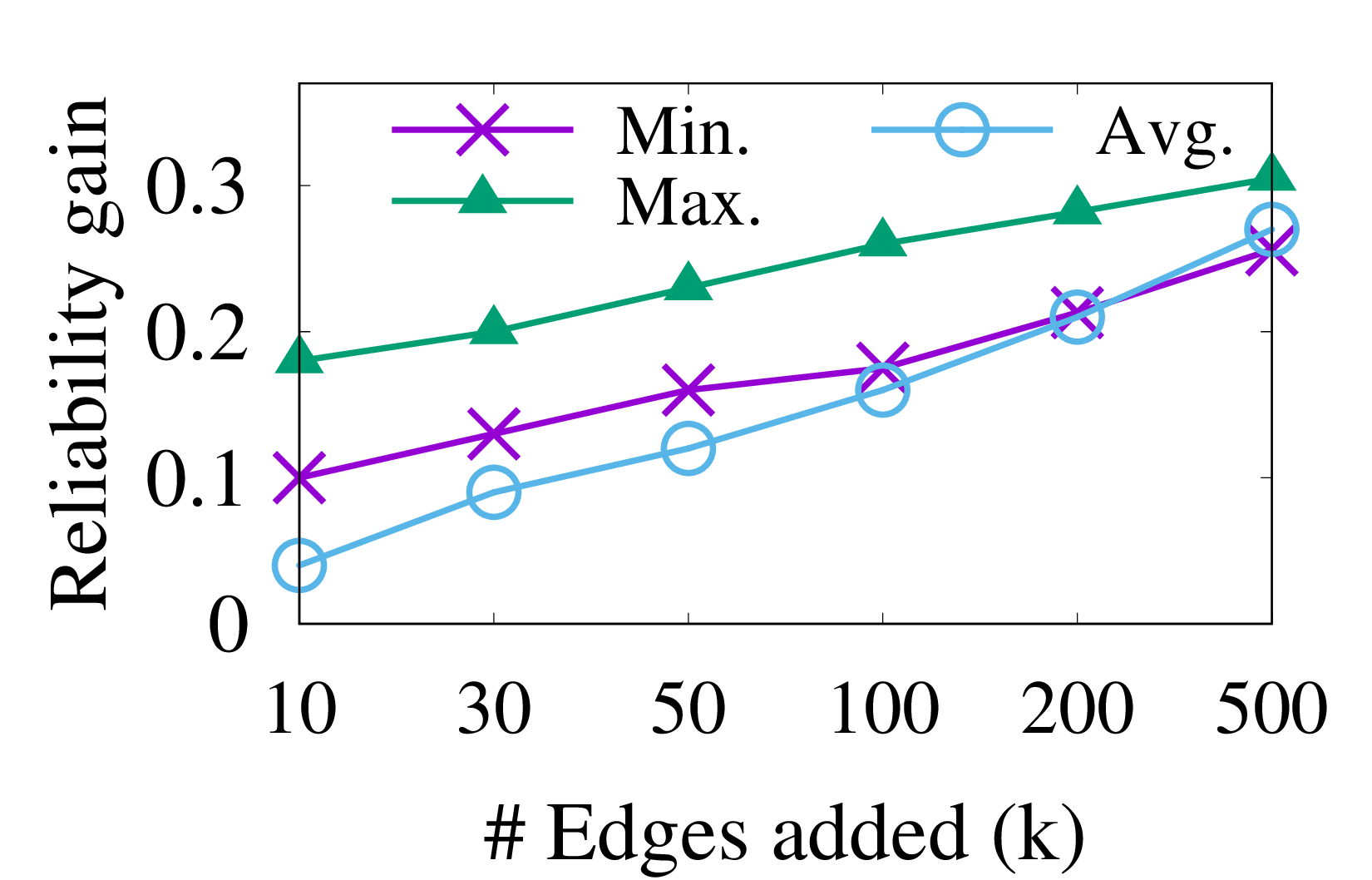}
		\label{fig:k_gain_mul}}
	\subfigure[\scriptsize {Running time, {\em Twitter}}]
	{\includegraphics[scale=0.146,angle=270]{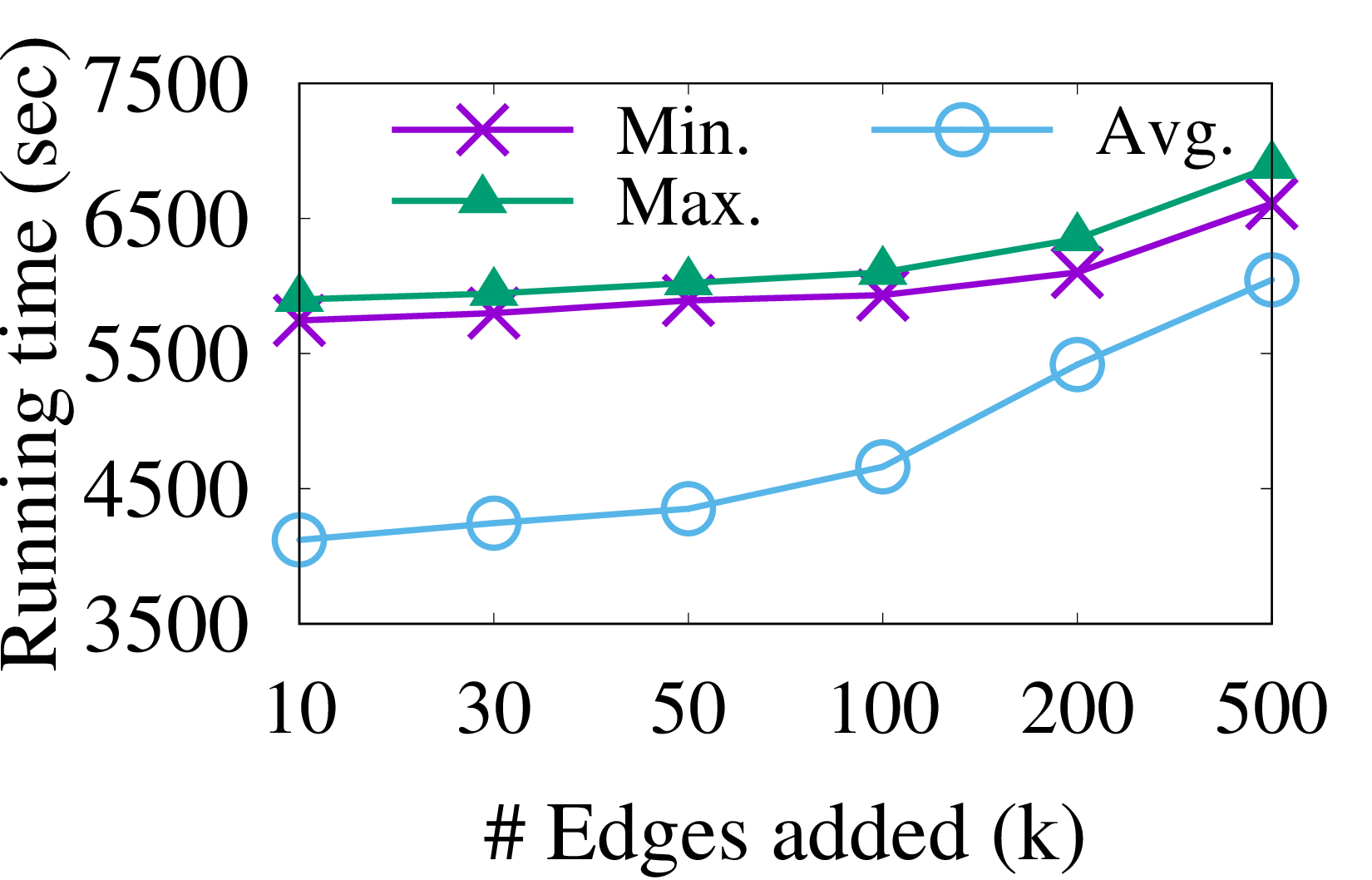}
		\label{fig:k_time_mul}}
	\vspace{-4mm}
	\caption{Reliability gain and running time comparison with varying budget on \#new edges $k$. $\zeta=0.5$, $r=100$ $l=30$, \#Sources=\#Targets=100, {\em Twitter}.}
	\label{fig:varyK_multi}
	\vspace{-6mm}
\end{figure}

\vspace{-1.1mm}
\spara{Varying the budget $k$ on \#new edges.} Similar to single-source-target case, we vary $k$, now in a larger scale: 10 to 500,
and present the result in Figure \ref{fig:varyK_multi}. The reliability gains for all three aggregate functions increase
with larger $k$. 
The running time of BE with {\em Minimum/Maximum} aggregate function is less sensitive to a larger $k$,
since the complexity of their top-$k$ edge selection part remains the same as single-source-target case,
while the search space elimination part scales up. On the contrary, the running time of BE with {\em Average}
is almost linear to $k$. However, {\em Average} is still less time consuming than {\em Minimum/Maximum} with large $k$.

\vspace{-2mm}
\subsection{Case Study and Application}
\label{sec:eff}
\vspace{-1mm}

We demonstrate the effectiveness of our proposed methods via a case study about improving the $s$-$t$ reliability in a sensor network, and an application of maximizing average multi-source-target reliability in influence maximization.

\vspace{-1mm}
\subsubsection{Case study in sensor network}
\vspace{-1mm}
\label{sec:case}

\begin{figure}[t!]
	\vspace{-1mm}
	\centering
	\includegraphics[scale=0.24,angle=0]{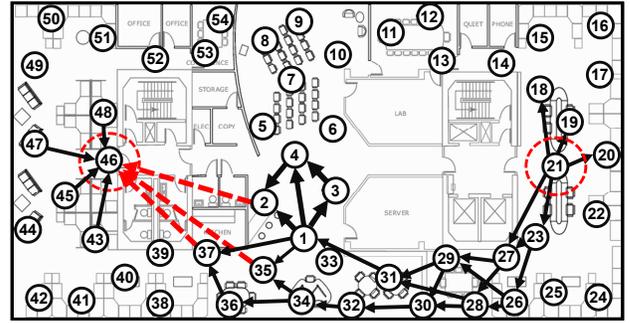}
	\vspace{-2.5mm}
	\caption{Improving the reliability from sensor 21 (right) to 46 (left) with 3 new links (marked by dotted lines).}
	\label{fig:sensor_1}
	\vspace{-3mm}
\end{figure}

\begin{figure}[t!]
	\centering
	\includegraphics[scale=0.24,angle=0]{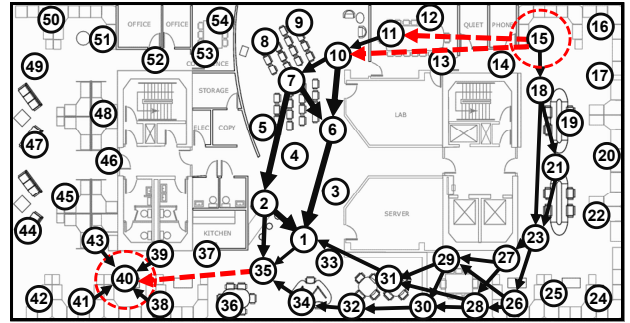}
	\vspace{-2.5mm}
	\caption{Improving the reliability from sensor 15 to 40 (on the diagonal) with 3 new links (marked by dotted lines).}
	\label{fig:sensor_2}
	\vspace{-5mm}
\end{figure}

To demonstrate the effectiveness of our problem, we conduct a case study on the {\em Intel Lab Data} (\url{http://db.csail.mit.edu/labdata/labdata.html}). This dataset contains the sensor network information with 54 sensors deployed in the Intel Berkeley Research Lab (map given in Figures \ref{fig:sensor_1} and \ref{fig:sensor_2}) between February 28th and April 5th, 2004.
The probabilities on links denote the percentages of messages from a sender successfully reached to a receiver.
The average link probability is 0.33 (ignoring edge probabilities which are lower than 0.1).

Assume that our goal is to maximize the reliability from: (1) a sensor on the
right hand side of the lab to a sensor on the left hand side (e.g., from sensor
21 to 46, with original reliability {\em 0.40}); (2) between two sensors on the diagonal of the lab (e.g., from
sensor 15 to 40, with original reliability {\em 0.28}). Due to budget constraints, only 3
new links are allowed for each case. We further assume that the probability of
each new link would be the same as the average edge probability of the original
dataset, that is, 0.33.  We notice that if two sensors are more than 20 meters
away, the original link probability between them is usually close to 0.  Thus,
we only allow establishing new links between a pair of sensors that are at most
15 meters away.

Figure \ref{fig:sensor_1} demonstrates the solution obtained by our algorithm for
case (1).  Only those links with probabilities higher than the average value
(0.33) are shown in the figure, and the thickness represents link
probabilities. Clearly, sensor 46 has very weak connections from outside, while
sensor 21 is connected with the bottom part of the lab. A very dense network
exists in the bottom part of the lab.  Therefore, our solution for this case is
to connect sensor 46 with sensors in the bottom part of the lab. By
establishing three links 2 to 46, 35 to 46, and 37 to 46, we improve the
reliability between 21 to 46 {\em from 0.40 to 0.88}.

For case (2), notice in Figure \ref{fig:sensor_2} that the sensors in the
center part of lab are well-connected, and the links in this region are thicker
than those in the bottom part. The source sensor 15 has a few
connections with sensors in the bottom part, but no link with those in
the center part. The destination sensor 40 has limited connections
beyond its physical neighbors. Existing configuration offers a poor reliability
of 0.28 for the connection between source and destination which we like to
improve. The smart decision made by our algorithm is as follows: First,
connect sensor 35 to 40, thus making sensor 35 a bridge between the
center and the bottom region of the network; Second, enable connection from
sensor 15 to the center part (by establishing link from 15 to 10, and from 15
to 11).  This results in {\em 0.58} overall reliability from sensor 15 to 40,
which is more than double of the original reliability value.  These results
illustrate how our proposed solution for the budgeted reliability maximization
problem can be useful in solving real-life problems.

\vspace{-2mm}
\subsubsection{Application in influence maximization}
\vspace{-1mm}
\label{sec:m_app}
In social influence maximization following the widely-used independent cascade model \cite{KKT03},
when some node $u$ first becomes active at step $t$, it gets a single chance to activate each of its currently inactive out-neighbors $v$ at step $t+1$, with probability $p(u,v)$.
Initially, only the source nodes are active, and the activation continues in discrete steps. When no more nodes can be activated, the number of active nodes in target set
is referred to as the influence spread. With possible world notation, the influence spread from source set $S$ to target set $T$ can be formulated as:
\vspace{-2mm}
\begin{align}
Inf(S,T)=\underset{G \sqsubseteq \mathcal{G}}{\sum}\left[Pr(G)\underset{t\in  T}{\sum}I_G(S,t)\right]
\vspace{-5mm}
\end{align}
As discussed in \S~\ref{sec:max_avg}, the {\em average reliability} from $S$ to $T$ is:
\vspace{-2mm}
\begin{align}
R_{avg}(S,T)=\frac{1}{|S||T|}\underset{G \sqsubseteq \mathcal{G}}{\sum}\left[Pr(G)\underset{s,t\in S\times T}{\sum}I_G(s,t)\right]
\vspace{-5mm}
\end{align}
Clearly, in each possible world, if we only check whether there is {\em at least one} path to $t$ from any $s\in S$, instead of counting the {\em exact number} of $s\in S$ which has a path to $t$,
our problem becomes equivalent to the (targeted) influence maximization problem. Adding a new edge in this network implies
recommending and/or establishing collaboration with an author in the real-world \cite{ChaoRRB12}.

\begin{figure}[t!]
	\centering
	\vspace{-1.5mm}
	\includegraphics[scale=0.19,angle=270]{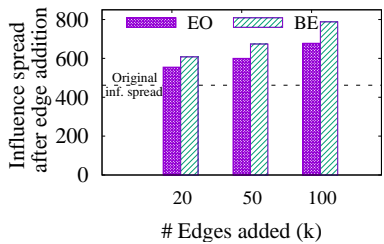}
	\vspace{-1mm}
	\caption{\small Influence spread comparison in {\em Databases} area. $\zeta=0.5$, $r=100$, $l=30$, $\frac{k_1}{k}=10\%$, {\em DBLP}.}
	\label{fig:inf}
	\vspace{-3.5mm}
\end{figure}

In {\em DBLP} dataset, we select a set of junior researchers in {\em Databases} area, containing 1000 authors
randomly selected from all the authors with 1-3 papers in [SIGMOD, VLDB, ICDE].
Similarly, we choose 50 senior researchers with more than 10 papers in [SIGMOD, VLDB, ICDE],
uniformly at random. The expected influence spread from the senior to the junior group is around {\em 462},
using the IC model. Next, we aim at maximally improving the influence spread from the senior group to the junior group,
by adding up to 100 new edges. As shown in Figure \ref{fig:inf}, our method 
outperforms Eigenvalue-based optimization (EO) \cite{CTPEFF16}, and results in about 326 more
influenced authors within the junior set.

\vspace{-3mm}
\section{Conclusions}
\vspace{-1mm}
In this paper, we introduced and investigated the novel and fundamental problem of maximizing the reliability between a given pair of nodes in an uncertain graph by adding
a small number of edges. We proved that this problem is \NP-hard and also hard to approximate.
Several interesting observations are presented to characterize it.
Our purposed solution first eliminates the search space based on original reliability, and
then selects the top-$k$ edges following an iterative most-reliable path-batches inclusion
algorithm. We further studied one restricted and several extended versions of the problem, to support a wider family of queries.
The experimental results validated the effectiveness, efficiency, and scalability of our method, and rich real-world case studies
demonstrated the usefulness of our budgeted reliability maximization problem. In future, a total reliability budget on new edges,
instead of a fixed/ individual budget on each new edge, can be considered. This will add more complexity on selecting
proper candidate edges and allocating reliability budget to them.

\vspace{-5mm}
{
	\bibliographystyle{IEEEtran}
	\bibliography{ref}
}

%

\IEEEoverridecommandlockouts
\vspace{-10mm}
\begin{IEEEbiography}[{\includegraphics[width=0.7in,height=0.9in]{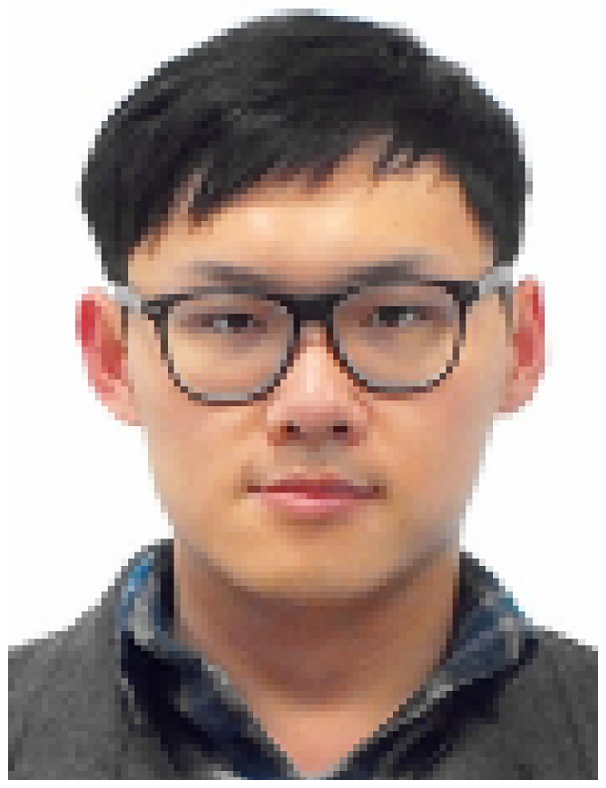}}]
{Xiangyu Ke} is a Ph.D. candidate in Nanyang Technological University, Singapore. He obtained his B.E. in Computer Science and Technology 
from Southeast University, China in 2016. His research interests include querying and mining of complex graphs. He published papers in SIGMOD, PVLDB, CIKM, and received SIGMOD 2018 Travel Award.
\end{IEEEbiography}
\vspace{-17mm}
\begin{IEEEbiography}[{\includegraphics[width=0.7in,height=0.9in]{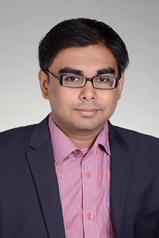}}]
{Arijit Khan} is an Assistant Professor at Nanyang Technological University, Singapore. He earned his PhD from UC Santa Barbara, and did a post-doc in ETH Zurich. 
He is the recipient of the IBM PhD Fellowship in 2012-13. He co-presented tutorials on graph queries and systems at ICDE 2012, VLDB 2014, 2015, 2017, and  wrote a book on uncertain
graphs in Morgan \& Claypool's Synthesis Lectures on Data Management. 
\end{IEEEbiography}
\vspace{-15mm}
\begin{IEEEbiography}[{\includegraphics[width=0.7in,height=0.9in]{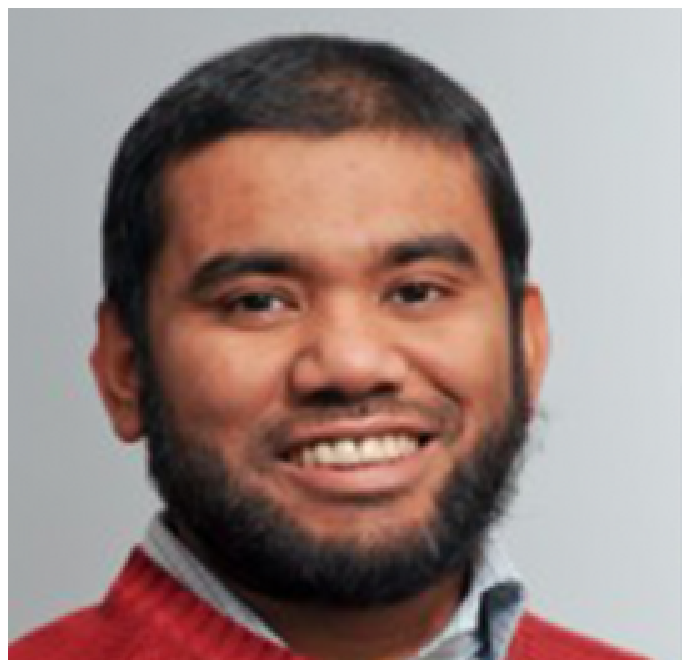}}]
{Mohammad Al Hasan} is an associate professor at Indiana University -- Purdue University, Indianapolis. Before that, he was a 
senior research scientist at eBay Research, San Jose. He earned his PhD from 
Rensselaer Polytechnic Institute, NY. He received the PAKDD best paper award (2009), SIGKDD doctoral dissertation award (2010), and NSF CAREER award (2012).
\end{IEEEbiography}
\vspace{-17mm}
\begin{IEEEbiography}[{\includegraphics[width=0.7in,height=0.9in]{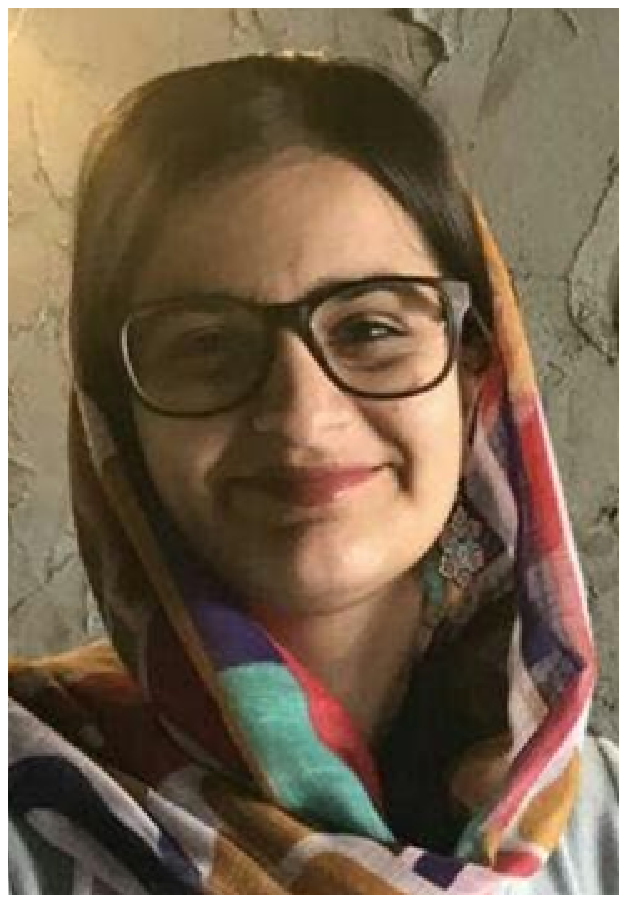}}]
{Rojin Rezvansangsari} is an undergraduate student in Sharif University of Technology. This work was done during her research internship in Nanyang Technological University, Singapore.
\end{IEEEbiography}

\vfill



\end{document}